\documentclass[twocolumn]{aastex62}

\usepackage{hyperref}
\usepackage{amsmath}

\received{October 3, 2022}
\revised{October XX, 2022}
\accepted{October YY, 2022}

\submitjournal{ApJ}

\shorttitle{The JWST Hubble Sequence}
\shortauthors{Ferreira et al.}

\def\solm{M$_{\odot}\,$}

\def\solm{M$_{\odot}\,$}

\def\casgm20{CAS-G-M$_{20}\,$}
\def\m20{M$_{20}\,$}

\usepackage{tikz}
\usetikzlibrary{shapes,arrows}

\tikzstyle{decision} = [diamond, draw, fill=white!20, 
    text width=7em, text badly centered, node distance=3cm, inner sep=0pt]
\tikzstyle{block} = [circle, draw, fill=blue!20, 
    text width=7em, text centered, rounded corners, minimum height=3em]
\tikzstyle{start} = [rectangle, draw, fill=black, 
    text width=7em, text centered, rounded corners, minimum height=3em]
\tikzstyle{final} = [rectangle, draw, fill=red!10, 
    text width=7em, text centered, minimum height=3em]
\tikzstyle{line} = [draw, -latex']
\tikzstyle{cloud} = [draw, ellipse,fill=red!20, node distance=3cm,
    minimum height=2em]

\begin{document}

\title{The JWST Hubble Sequence: The Rest-Frame Optical Evolution of Galaxy Structure at $1.5 < z < 8$} 

\correspondingauthor{Leonardo Ferreira}
\email{leonardo.ferreira@nottingham.ac.uk}

\author[0000-0003-1949-7638]{Leonardo Ferreira}
\affil{Centre for Astronomy and Particle Theory, University of Nottingham,
Nottingham, UK}

\author[0000-0003-1949-7638]{Christopher J. Conselice}
\affiliation{Jodrell Bank Centre for Astrophysics, University of Manchester, Oxford Road, Manchester UK}

\author[0000-0001-6245-5121]{Elizaveta Sazonova}
\affiliation{Department of Physics and Astronomy, Johns Hopkins University, Baltimore, MD 21218, USA}

\author[0000-0002-0056-1970]{Fabricio Ferrari}
\affiliation{Instituto de Matemática Estatística e Física, Universidade Federal do Rio Grande, Rio Grande, RS, Brazil}

\author[0000-0002-6089-0768]{Joseph Caruana}
\affiliation{Department of Physics, University of Malta, Msida MSD 2080}
\affiliation{Institute of Space Sciences \& Astronomy, University of Malta, Msida MSD 2080}

\author[0000-0003-2527-0819]{Clár-Bríd Tohill}
\affil{Centre for Astronomy and Particle Theory, University of Nottingham, Nottingham, UK}

\author[0000-0002-2410-1776]{Geferson Lucatelli}
\affiliation{Jodrell Bank Centre for Astrophysics, University of Manchester, Oxford Road, Manchester UK}

\author[0000-0003-4875-6272]{Nathan Adams}
\affiliation{Jodrell Bank Centre for Astrophysics, University of Manchester, Oxford Road, Manchester UK}

\author[0000-0003-2946-8080]{Dimitrios Irodotou}
\affiliation{Department of Physics, University of Helsinki, Gustaf Hällströmin katu 2, FI-00014, Helsinki, Finland}

\author[0000-0001-6434-7845]{Madeline A. Marshall} 
\affiliation{National Research Council of Canada, Herzberg Astronomy \&
Astrophysics Research Centre, 5071 West Saanich Road, Victoria, BC V9E 2E7,
Canada}
\affiliation{ARC Centre of Excellence for All Sky Astrophysics in 3 Dimensions
(ASTRO 3D), Australia}

\author[0000-0002-3257-8806]{Will J. Roper}
\affiliation{Astronomy Centre, University of Sussex, Falmer, Brighton BN1 9QH, UK}

\author[0000-0001-7964-5933]{Christopher C. Lovell}
\affiliation{Centre for Astrophysics Research,  School of Physics, Engineering \& Computer Science, University of Hertfordshire, Hatfield AL10 9AB, UK}

\author[0000-0002-0730-0781]{Aprajita Verma}
\affiliation{Sub-department of Astrophysics, University of Oxford, Denys Wilkinson Building, Keble Road, Oxford, OX1 3RH.}

\author[0000-0003-0519-9445]{Duncan Austin}
\affiliation{Jodrell Bank Centre for Astrophysics, University of Manchester, Oxford Road, Manchester, UK}

\author[0000-0002-9081-2111]{James Trussler}
\affiliation{Jodrell Bank Centre for Astrophysics, University of Manchester, Oxford Road, Manchester UK}

\author[0000-0003-3903-6935]{Stephen M.~Wilkins} %
\affiliation{Astronomy Centre, University of Sussex, Falmer, Brighton BN1 9QH, UK}
\affiliation{Institute of Space Sciences and Astronomy, University of Malta, Msida MSD 2080, Malta}



\begin{abstract}

We present results on the morphological and structural evolution of a total of 4265 galaxies observed with JWST at $1.5 < z < 8$ in the JWST CEERS observations that overlap with the CANDELS EGS field. This is the biggest visually classified sample observed with JWST yet, $\sim20$ times larger than previous studies, and allows us to examine in detail how galaxy structure has changed over this critical epoch. All sources were classified by six individual classifiers using a simple classification scheme aimed to produce disk/spheroid/peculiar classifications, whereby we determine how the relative number of these morphologies evolves since the Universe's first billion years. Additionally, we explore structural and quantitative morphology measurements using \textsc{Morfometryka}, and show that galaxies at $z > 3$ are not dominated by irregular and peculiar structures, either visually or quantitatively, as previously thought. We find a strong dominance of morphologically selected disk galaxies up to $z = 8$, a far higher redshift than previously thought possible.  We also find that the stellar mass and star formation rate densities are dominated by disk galaxies up to $z \sim 6$, demonstrating that most stars in the universe were likely formed in a disk galaxy.   We compare our results to theory to show that the fraction of types we find is predicted by cosmological simulations, and that the Hubble Sequence was already in place as early as one billion years after the Big Bang. Additionally, we make our visual classifications public for the community.
\end{abstract}

\keywords{galaxy morphology, high redshift galaxies, JWST, visual classifications, galaxy formation}

\vspace{1em}
\section{Introduction} \label{sec:intro}

Since the discovery of galaxies, a principal aim of their study has been to characterize their structures and morphologies. The very fact that galaxies appear to be extended, as opposed to point sources, already provides an elusive clue to their nature being different from that of the stars. In fact, it can be said that it was the extended nature of these objects that instigated the debate about whether they were external to our own galaxy, a problem solved through obtaining distances to these systems \citep{Hubble1926}.  Even before then, however, the fact that the structure of galaxies holds important information had been known since at least the time of Lord Rosse and his discovery of spiral structure in nearby massive galaxies such as M51 \citep{Rosse1850}.

Since that time, galaxy structure, morphology, and how these properties evolve with time has remained a key aspect to understanding galaxy evolution \citep[e.g.,][]{Conselice2003a,Lotz2004, Mortlock2013, Delgado-Serrano2010, Schawinski2014, Conselice2014a, Whitney2021,FERREIRA2022b}. The resolved structure of distant galaxies, in particular with the advent of the Hubble Space Telescope, clearly revealed that faint distant galaxies were more peculiar and irregular, and few fit into the Hubble sequence \citep[e.g.][]{Driver1995}.  Later, once redshifts became available, it became clear that galaxy structures evolve strongly, but systematically, with redshift, such that peculiar galaxies dominate the population at $z > 2.5$ \citep[e.g.,][]{Conselice2008, Conselice2014a, Whitney2021}.  However, because of the limited red wavelengths of Hubble, we still have not yet been able to trace the rest-frame optical light of galaxies back to within the first few Gyr of the Big Bang.  The F160W band on the Hubble Space Telescope (HST) can only probe rest-frame visible light up to $z \sim 2.8$, but JWST permits us to obtain the same type of data out to $z\sim9$ with F444W \citep[e.g.,][]{FERREIRA2022b}. Moreover, JWST's superior resolution and longer wavelength filter set allows galaxy structure to be better measured than with the lower resolution of HST.

Observations of galaxy structure and morphology at $z > 3$ do show that in the rest-frame UV galaxies are peculiar and irregular \citep[e.g.,][]{conselice2009}. Moreover, galaxies are often very clumpy at these redshifts, as seen with deep WFC3 data \citep[e.g., ][]{Oesch2010}.  Furthermore, observations of pairs of galaxies show that the merger rate and fraction of galaxies up to $z \sim 6$ is very high, and therefore that galaxy structure should likewise be affected significantly \citep[e.g.,][]{duncan2019}.  At the same time, we know that the Hubble sequence is established at $z < 1$ \citep[e.g.,][]{Mortlock2013}. However, whether the Hubble sequence already existed in the earlier Universe remains an open question. 

While earlier HST-based studies found that the dominating majority of galaxies at $z>2$ are peculiar, recent JWST-based studies find a high number of regular disk galaxies at high redshift  \citep[e.g.,][]{FERREIRA2022b, NELSON2022, Jacobs2022, MORFEUS2022}, consistent with an even earlier emergence of the Hubble sequence.   The nature of this evolution and its implications are however still unknown.

Quantitative measures of galaxy structure and morphology also present stringent constraints for numerical simulations to reproduce. In recent years, full hydrodynamic simulations \citep{Schaye2015, Nelson2019, Lovell2021, Madeline2022} enable resolved morphologies to be predicted in a self-consistent manner, and recent novel simulation approaches allow these to be tested out to the highest redshifts \citep{Roper2022}. There are a number of difficulties when comparing morphologies between simulations and observations, however simple measures of the abundance of e.g. disk and elliptical galaxies can provide hints as to the underlying mechanisms leading to morphological evolution.  However, what we known from early JWST work is that the morphological and structural features of galaxies at $z > 1$ are much different than what was found with HST \citep[e.g.,][]{Ferreira2020}, and therefore a more thorough analysis is needed to address these fundamental problems.

Thus, in this paper we explore the morphological properties of 4265 galaxies observed with JWST through visual galaxy classifications and quantitative morphology, from $z=1.5$ to 8. This is a sample 20 times larger than any previous morphological and structural study using JWST.  Amongst other things, we confirm that these early galaxies have predominantly disk morphologies, and that the Hubble sequence appars to be already established as early as $z\sim8$.  In this paper we discuss these results and their implications for galaxy formation and evolution.

The paper is organized as follows. In \S~\ref{sec:data} we describe the data products used, our reduction pipeline, the visual classification scheme adopted, as well as our methods of quantitative morphology. \S~\ref{sec:results} describes the results from our classification effort, quantitative morphology measurements including a discussion on the evolution of the Hubble sequence from $z=1.5$ to 8. We finish with a summary of our main results in \S~\ref{sec:summary}.

\section{Data and Methods}\label{sec:data}

We use the public NIRCam JWST observations from the Cosmic Evolution Early Release Science Survey (CEERS; PI: Finkelstein, ID=1345, Finkelstein et al. in prep), that overlap with the Cosmic Assembly Near-IR Deep Extragalactic Legacy Survey (CANDELS; \citealt{Grogin2011, Koekemoer2011}) on the Extended Growth Strip field (EGS). These data are reduced independently using a custom set-up of the \textsc{JWST} pipeline version \textsc{1.6.2} using the on-flight calibration files available through the \textsc{CDRS 0942}, an extensive description is given in \S~\ref{sec:reduction}. 
    
We select 4265 sources with $z > 1.5$ from the CANDELS catalogs which overlap with the area covered by CEERS. We take advantage of the robust photometric redshifts, star formation rates and stellar masses already derived for CANDELS in previous works \citep{Duncan2014, duncan2019, Whitney2021} to conduct this analysis. Neither morphological information is used for the selection of sources, nor are magnitude cuts employed, as we want to make sure that we include sources that might be faint in HST, but bright in JWST observations. This is also the case for morphology; we are also interested in sources that can show dramatic changes in morphology between the two instruments. 

We employ two different approaches to these data: first we perform visual classifications for all sources, which is described in detail in \S \ref{sec:visual}. Second, we perform quantitative morphology through \textsc{Morfometryka} \citep{ferrari2015}, where we measure non-parametric morphology estimates such as CAS, G-M20, H, Spirality, Sizes, as well as light profile fitting, which is described in detail \S \ref{sec:mfmtk}. 

\subsection{Data Reduction}\label{sec:reduction}

We reprocess all of the uncalibrated lower-level JWST data products for this field following our modified version of the JWST official pipeline. This is similar to the process used in \citet{adams2022} and \citet{FERREIRA2022b}, but with minor updates and improvements, and can be summarised as follows: (1) We use version 1.6.2 of the pipeline with the Calibration Reference Data System (CRDS) version 0942 which was the most up-to-date version at the time of writing. Use of CRDS 0942 is essential for zero point issues we discuss in \citep{adams2022}.  (2) We apply the 1/f noise correction derived by Chris Willott on the resulting level 2 data of the JWST pipeline.\footnote{\url{https://github.com/chriswillott/jwst}} (3) We extract the sky subtraction step from stage 3 of the pipeline and run it independently on each NIRCam frame, allowing for quicker assessment of the background subtraction performance and fine-tuning. (4) We align calibrated imaging for each individual exposure to GAIA using \texttt{tweakreg}, part of the DrizzlePac python package.\footnote{\url{https://github.com/spacetelescope/drizzlepac}} (5) We pixel-match the final mosaics with the use of \texttt{astropy reproject}.\footnote{\url{https://reproject.readthedocs.io/en/stable/}} The final resolution of the drizzled images is 0.03 arcseconds/pixel. There is rapid development in the above procedure, and so we anticipate future studies to continue to make refinements to the JWST pipeline. Each one of the four June CEERS observations was processed into individual mosaics.

\subsection{Photometric Redshifts and Stellar Masses}

The photometric redshifts that we use in this paper originate from the redshifts calculated in \citet{duncan2019} for EGS.  These are based on the original CANDELS+GOODS WFC3/ACS imaging and data, Spitzer/IRAC S-CANDELS \citep{Ashby2015}, and ground-based observations with CFHT \citep{Stefanon2017}.  The overall method for this is described in detail in \citet{duncan2019}.

\subsection{Visual Classification}\label{sec:visual}

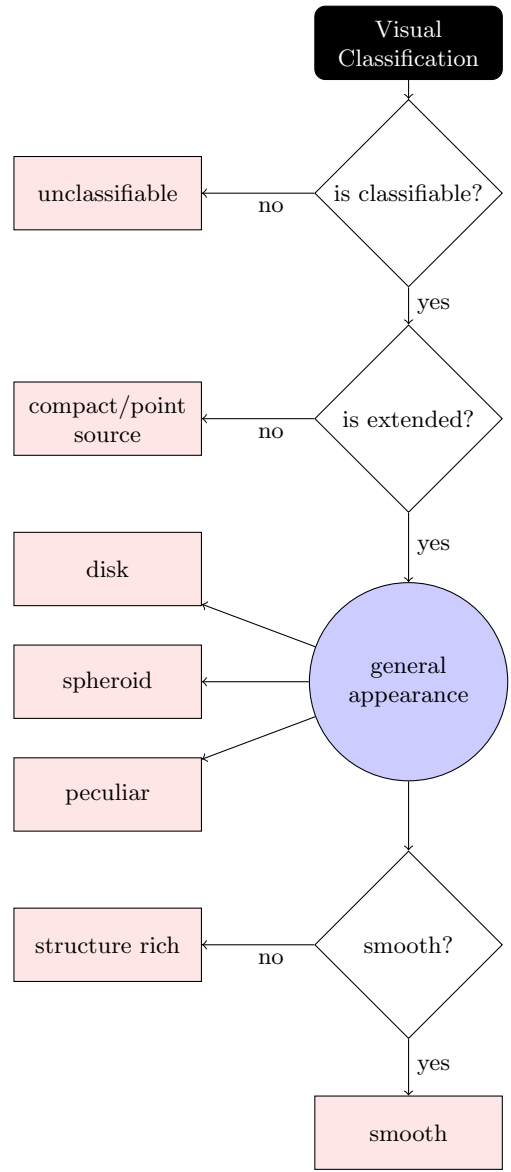
\begin{figure}
\centering
\begin{tikzpicture}[auto]
    \node [start] (begin) {\textcolor{white}{Visual \\ Classification}};
    \node [decision, below of=begin, node distance=2.0cm] (classifiable) {is classifiable?};
    \node [final, left of=classifiable, node distance=4cm] (unclassifiable) {unclassifiable};
    \draw[->]      (begin) --  (classifiable);

    \node [decision, below of=classifiable] (extended) {is extended?};
    
    \node [final, left of=extended, node distance=4cm] (compact) {compact/point source};
    \draw[->]      (extended) -- node[text width=0cm] {no} (compact);
    \node [block, below of=extended, node distance=3.5cm] (appearance) {general appearance};

    \node [final, left of=appearance, node distance=4cm] (spheroid) {spheroid};
    \node [final, above of=spheroid, node distance=1.5cm] (disk) {disk};
    \node [final, below of=spheroid, node distance=1.5cm] (peculiar) {peculiar};
    \draw[->]      (extended) -- node[text width=0cm] {yes} (appearance);
    \draw[->]      (classifiable) -- node[text width=0cm] {yes} (extended);
    \draw[->]      (classifiable) -- node[text width=0cm] {no} (unclassifiable);
    \draw[->]      (appearance) --  (disk);
    \draw[->]      (appearance) --  (spheroid);
    \draw[->]      (appearance) --  (peculiar);
    
     \node [decision, below of=appearance, node distance=3.5cm] (smooth) {smooth?};
     \node [final, below of=smooth, node distance=2.5cm] (issmooth) {smooth};
      \node [final, left of=smooth, node distance=4cm] (structure) {structure rich};
      
      \draw[->]      (appearance) --  (smooth);
      \draw[->]      (smooth) -- node[text width=0cm] {yes}(issmooth); 
      \draw[->]      (smooth) -- node[text width=0cm] {no} (structure);
\end{tikzpicture}
\caption{Fluxogram of our visual classification process. The classification is based on 4 basic questions that can produce a simple disk/spheroid/peculiar/compact classification and additional flags regarding whether the source is smooth or structurally rich.}
\label{fig:fluxogram}
\end{figure}

As a way to define the morphologies of the galaxies in our sample of 4265 sources, we construct a simple classification scheme that yields a large amount of information with a small number of classification questions, as opposed to having a very detailed sub-classification scheme of structure sub-components. The classification scheme is summarized in the fluxogram in Fig.~\ref{fig:fluxogram}. At high redshift, fine structural details are often difficult to recover and in general are ambiguous, hence these questions capture the overall appearance of the source. 

Our sample is the biggest visually classified sample observed with JWST yet, $\sim20$ times larger than what is reported in previous JWST morphology results \citep{FERREIRA2022b, NELSON2022, Jacobs2022}. A brief description of each possible resulting class is given below:

\begin{itemize}
\item Unclassifiable: Galaxies not clearly visible, too faint/noisy, and image issues like artifacts, cosmic rays. 
\item Point Sources: Sources that are smaller in angular size than the $\rm PSF_{FWHM}$ or that present clear wings/spike patterns consistent with point-like objects but no extended component.
\item Disks: Galaxies with a resolved disk in the form of an outer area of lower surface brightness with a regularly increasing brightness towards the center of the galaxy. This classification does not depend on there being a feature present in the disk such as a spiral pattern or a bar, although one can be present. 
\item Spheroid: Resolved, symmetrically and centrally concentrated, with a smooth profile and are round/elliptical in shape. 
\item Peculiar: Well-resolved galaxies with a morphology which is dominated by a disturbance or peculiarity, where the disturbance dominates any smooth components. 
\end{itemize}

Each galaxy is further classified as smooth or structured, where structured galaxies have features standing out from the smooth stellar envelope, such as star-formation clumps, tidal features, and merger signatures. Galaxies with distinct disk and bulge components were also classified as structured. Finally, the classifiers were able to provide additional notes on each source to aid in future analysis. 

Ultimately, after all classifications were aggregated we determined the final class of each object as the the one receiving the majority of the votes, as discussed in detail in Sec. \ref{sec:visual}. In the cases where classifiers disagreed, we included an \texttt{ambiguous} class.

Based on this classification scheme (Fig.~\ref{fig:fluxogram}), six authors of this study (CBT, CC, ES, JC, GL, LF) classified all the 4265 sources. This effort produced a robust catalog where every galaxy has all six classifications combined in classification fractions, one for each individual question present in the scheme (Fig.~\ref{fig:fluxogram}). 

To perform each classification, the classifiers were given access to a web application built with \texttt{flask}, \texttt{jinja} and \texttt{bootstrap} specifically tailored for this task. The volunteers were presented with the rest-frame image in the filter that corresponds to the source redshift (minimizing for  $\lambda_{rest}/(1+z)$), an RGB image (F277W+F356W+F444W) generated with \textsc{Trilogy} \citep{Coe2015}, the PSF image of the respective filter together with the size of the $PSF_{FWHM}$, and a questionnaire that reproduces Fig.~\ref{fig:fluxogram}. Results are stored in a \texttt{MySQL} database, that is then reduced and aggregated with \texttt{pandas} \citep{pandas}.

Each individual classifier's results are combined in a single table using the following criteria. 

First, we define how many votes there are for each source - that is, the votes that are not considered to be \texttt{unclassifiable} and \texttt{point source}. Then, if at least $50\%$ of all votes are assigned to any of these individual classes (i.e.~3 or more votes are exclusively in these categories), we consider the source to be unclassifiable or point-like, by the labels \texttt{n/a} and \texttt{ps}, respectively. For the point-like sources, we compare its size to the $PSF_{\rm FWHM}$. If it is larger than the $PSF_{\rm  FWHM}$, we change its classification to \texttt{spheroid}. The same is done the other way around: sources smaller than the $PSF_{\rm FWHM}$ are changed programatically to \texttt{ps}.

Second, for all the rest of the sources that have more than $50\%$ of the votes in disk, spheroid or peculiar categories, we average each individual classification decision in a class fraction. Hence, this class fraction is only based on the number of good votes (i.e.~votes that are for classifiable and extended galaxies).

Third, to all galaxies that have a clear majority as $frac > 0.5$, we assign the given class as the final class. For all the remaining cases that the classifiers disagree on (e.g., 2 votes in each category), we define those sources to have an \texttt{ambiguous} class. 

Finally, we include a structure index, the \texttt{smooth\_fraction},  that is independent of the general appearance, designed as a large umbrella to capture sources with rich structures, sub-components, and merging features.

This framework enables sources that might be ambiguous between two classifiers to have a more robust classification. The class fractions can also be used to control the purity of the samples, as higher agreement will represent a less contaminated dataset. We proceed, however, with the final classifications assigned by the majority of the classifiers.

\subsection{\sc Morfometryka}\label{sec:mfmtk}

\textsc{Morfometryka}  \citep{ferrari2015} performs several structural measurements on galaxy images in a straightforward,  non-interactive way. It measures non-parametric morphometric quantities along with 1D and 2D single Sérsic model fitting \citep{Sersic1963}. The inputs are the galaxy and PSF images, from which it estimates the background with an iterative algorithm, segments the sources and defines the target. If desired, it filters out external sources using GalClean\footnote{\url{https://github.com/astroferreira/galclean}} \citep{Ferreira2018}. From the segmented region it calculates standard  geometrical  parameters (e.g., center, PA, axial ratio) using image moments; it performs photometry measuring fluxes in similar ellipses with the aforementioned parameters; point sources are masked with a sigma clipping criteria. From the luminosity growth curve it establishes the Petrosian radius and the Petrosian Region \citep{Petrosian1976}, inside which all the measurements will be made. The 1D Sérsic fit is performed on the luminosity profile and used  as inputs for a 2D Sérsic fit done with the galaxy and the PSF images. Finally, it measures several morphometric parameters \citep[concentrations; asymmetries; Gini; M20; entropy, spirality, curvature, among others, ][]{Abraham1994, Bershady2000, Conselice2003, Lotz2004, Conselice2008, Lotz2008, ferrari2015}. 

We run \textsc{Morfometryka} for all filters available but only report results for the band that matches closest the rest-frame optical of the source ($\lambda = 0.5 \mu m $-$ 0.7 \mu m$).

\subsection{BlueTides high-z Mocks}\label{sec:bluetides}
By way of a comparison of observations with theory, we also consider mock JWST images from the BlueTides Mock Image Catalogue \citep{Madeline2022,MockCatalogue}. This is a catalogue of mock image stamps of $\sim 100,000$ galaxies from $z=7$ to 12, with the particle distributions and SEDs of each galaxy taken from the BlueTides hydrodynamical simulation \citep{Feng2015}. 
The images are created with the NIRCam transmission curves and convolved with JWST model PSFs from WebbPSF \citep{Perrin2015} to produce a realistic mock image for each galaxy. The BlueTides mock images have a pixel scale of $0\farcs0155$ for the NIRCam short-wavelength filters, which is half our observed pixel scale, and $0\farcs0315$ for the long-wavelength filters, which is equivalent to our observed pixel scale. Here we use the F444W filter, which is the same filter that we use to probe the rest-frame optical of the real observations at $z > 7$.

It is important to note that the images of the BlueTides galaxies are created in the `face-on' direction, which is defined by the angular momentum of particles in the galaxy \citep[see][]{Marshall2020}. However, studies suggest that the angular momentum of early galaxies does not correlate with their morphological structure \citep[e.g.][]{Park2022}; indeed, visual inspection of the BlueTides images shows that this direction does not necessarily correspond to the visual morphological `face-on' direction. Thus, while we do not expect this feature of the simulation to highly affect our comparison, some bias could be present.

We select all galaxies that overlap with the redshift range probed here from $z=7$ to 8. For each source stamp we add Gaussian noise to match the depth of the CEERS observations levels. Then, we run {\sc Morfometryka} for all stamps together with the PSF generated with {\sc WebbPSF}.

\section{Results}\label{sec:results}

\begin{figure*}
    \centering
    \includegraphics[width=1\textwidth]{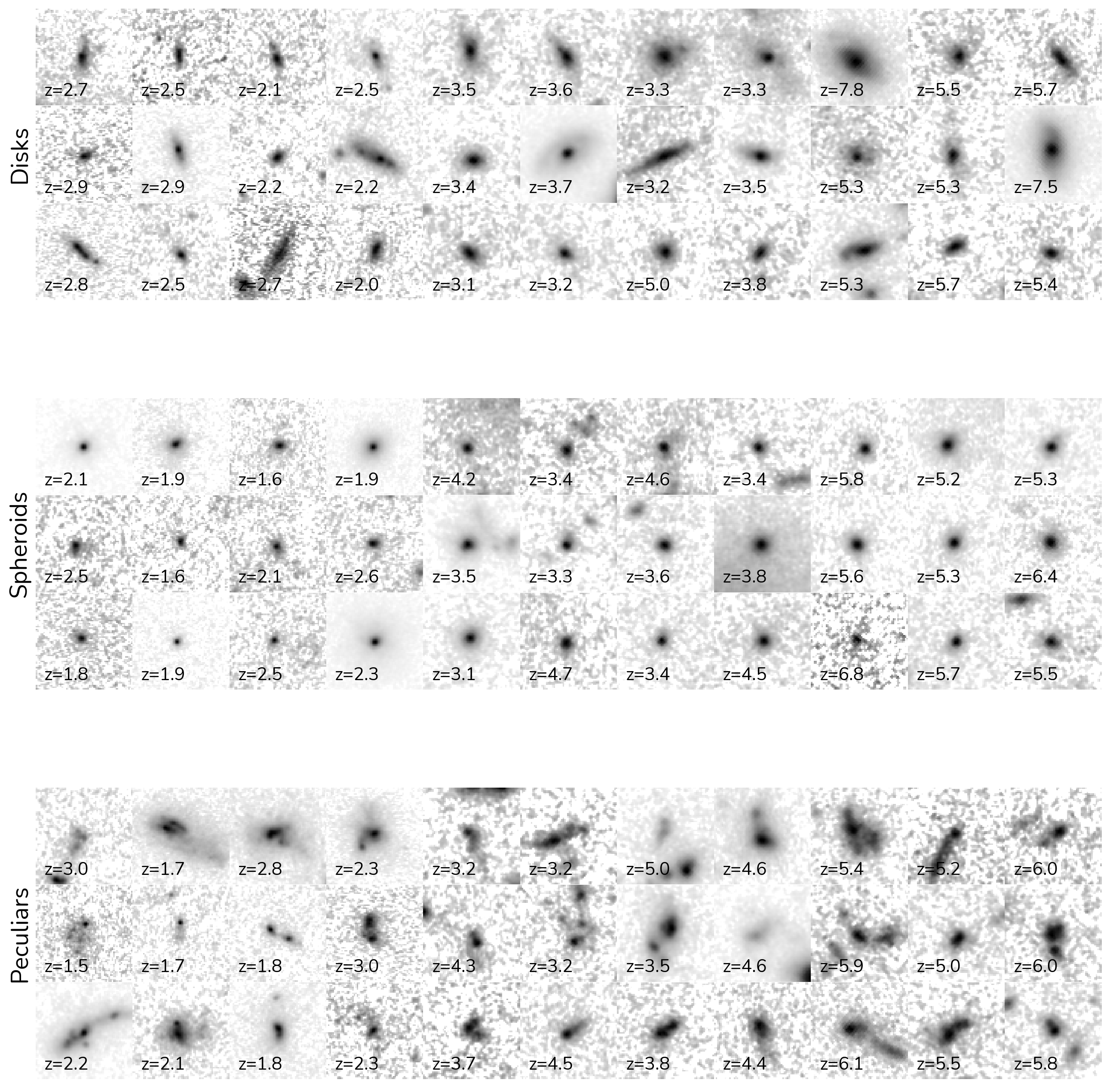}
    \caption{Rest-frame optical images for sources in our sample. The three panels show the three main classes \texttt{disks}, \texttt{spheroids} and \texttt{peculiars}, respectively. Galaxies are ordered horizontally by redshift, lowest redshifts in the left, highest redshifts to the right. Stamps are shown in square root normalization. Redshifts are from \cite{duncan2019} based on the CANDELS fields.}
    \label{fig:mosaic}
\end{figure*}

We report the morphology and structure evolution of the sample of 4265 galaxies (\S~\ref{sec:data}) based on visual classifications (\S~\ref{sec:visual}) and in quantitative morphology measurements (\S~\ref{sec:mfmtk}). The aggregated classifications catalog based on 6 independent classifiers contains 1696 disks ($\sim 40\%$), 561 spheroids ($\sim 13\%$), 1112 peculiars ($\sim 26\%$), 434 ambiguous sources ($\sim 10\%$), 66 point sources ($\sim 1\%$) and 396 unclassifiable sources ($\sim 9\%$). Examples of each of these types are shown in Figure~\ref{fig:mosaic} in bins of increasing redshift. The full catalog is publicly available\footnote{Full catalog will be made publicly available once this paper goes to press at \url{https://github.com/astroferreira/CEERS_JUNE_VISUAL_CLASSIFICATIONS}}.

These visual classifications are the basis for the discussion in this section. In Section \S~\ref{subsec:mainclasses} we detail the three base classes and the caveats from the visual classifications. We follow with a description of the quantitative morphologies of these sources and how they relate to the visual classifications in Section \S~\ref{subsec:mfmtk_evo}. We explore the evolution of the Hubble Sequence in \S~\ref{subsec:hubblesequence}, the evolution of the contribution of each morphological class to star formation and stellar mass in \S~\ref{subsec:starformationevo}. We compare these classifications with predictions from cosmological simulations in Section \S~\ref{subsec:simpreds}, and in \S~\ref{subsec:hstvjwst} we briefly discuss the main differences between HST and JWST imaging that could explain some of the discrepancies from previous studies.

\subsection{Disks, Spheroids and Peculiars}\label{subsec:mainclasses}

Figure~\ref{fig:mosaic} displays examples randomly drawn from the catalog for the three main morphological classes: disks, spheroids and peculiars, respectively. 

The visual distinction between these classes is clear, with the disks often showing two structural components in the form of a concentrated bulge and a disky envelope, while the spheroids are mostly single-profile, centrally concentrated sources, with some exhibiting PSF-like structure due to the central concentration or AGN emission. However, we note that for most cases, telling apart two types of light concentrations by eye is a difficult task, as sources at high redshifts do not show other clear, disk-like features such as spiral arms, bars and rings, and overall display lower concentrations \citep{Buitrago2008, Buitrago2013}. For a better distinction between face-on disks and spheroids, a quantitative approach such as S\'ersic fitting might be used alongside visual classification.

\begin{figure}
    \centering
    \includegraphics[width=0.35\textwidth]{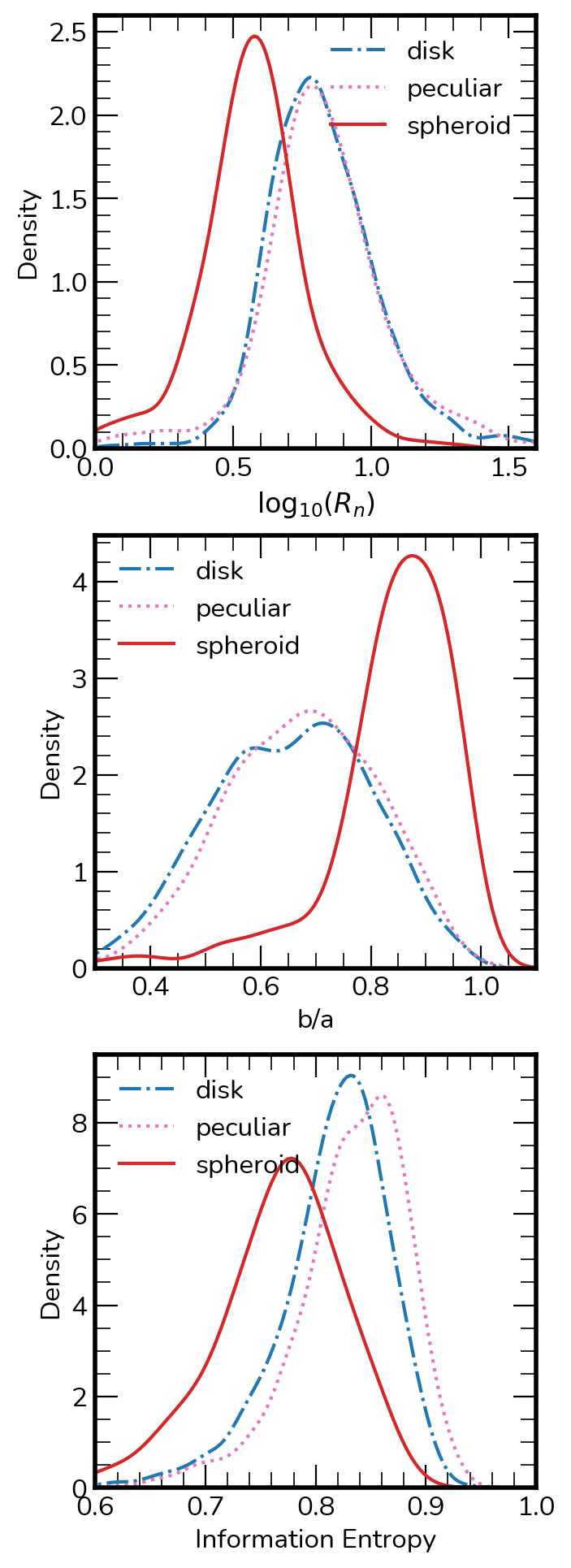}
    \caption{Effective radius ($R_n$) (top), axis ratio ($b/a$) (middle) and information entropy ($H$) (bottom). We show key measurements that clue to differences most used by the classifiers for the spheroid $\times$ other classes. Spheroids are defined by their lack of structure, low elongation and small sizes in general.}
    \label{fig:samplestats}
\end{figure}

Figure~\ref{fig:samplestats} shows three indicators for visual distinctions between the overall sample of {\tt disks} and {\tt spheroids}. {\tt spheroids} are more compact with lower effective radius, higher axis ratios, and lower information entropy (e.g., Shannon entropy) indicating lack of structure. The information entropy describes how the pixel values in an image are distributed, smooth galaxies will present low entropy while clumpy galaxies will have high entropy (see \citet{ferrari2015} for details on how it is calculated). These distributions follow what is found for high redshift spheroids in previous studies, in that they are round and smaller than disks \citep{Buitrago2013}. However, the axis ratios found here are on the high end, with a lack of spheroids with intermediary axis ratios $0.4 < b/a\sim0.7$. Ultimately, some biases might be present, such as the elongation/axis ratio causing some contamination, and thus we possibly miss some elongated spheroids, but this is expected due to each classifier's subjective perspective on what defines these. We advise the user of the catalog to leverage the class fractions to control purity by only selecting sources with strong agreement.

The peculiars on the other hand vary wildly, from mild disturbances, to clear signs of galaxy merging, often with a companion nearby. Additionally, some high redshift disks and spheroids might end up being classified as peculiars due to more asymmetric/disturbed morphologies than low redshift counterparts. However, as discussed in \S~\ref{subsec:mfmtk_evo}, for most peculiars, the quantitative morphology is consistent with disturbed morphologies.

\subsection{Quantitative Morphology Evolution}\label{subsec:mfmtk_evo}

Cross-examining the morphologies defined by eye using quantitative methods is essential for understanding how their appearance changes across cosmic time. We explore the visual morphologies with several quantitative morphology indicators, both non-parametric and parametric (\S~\ref{sec:mfmtk}).

\begin{figure*}
    \centering
    \includegraphics[width=\textwidth]{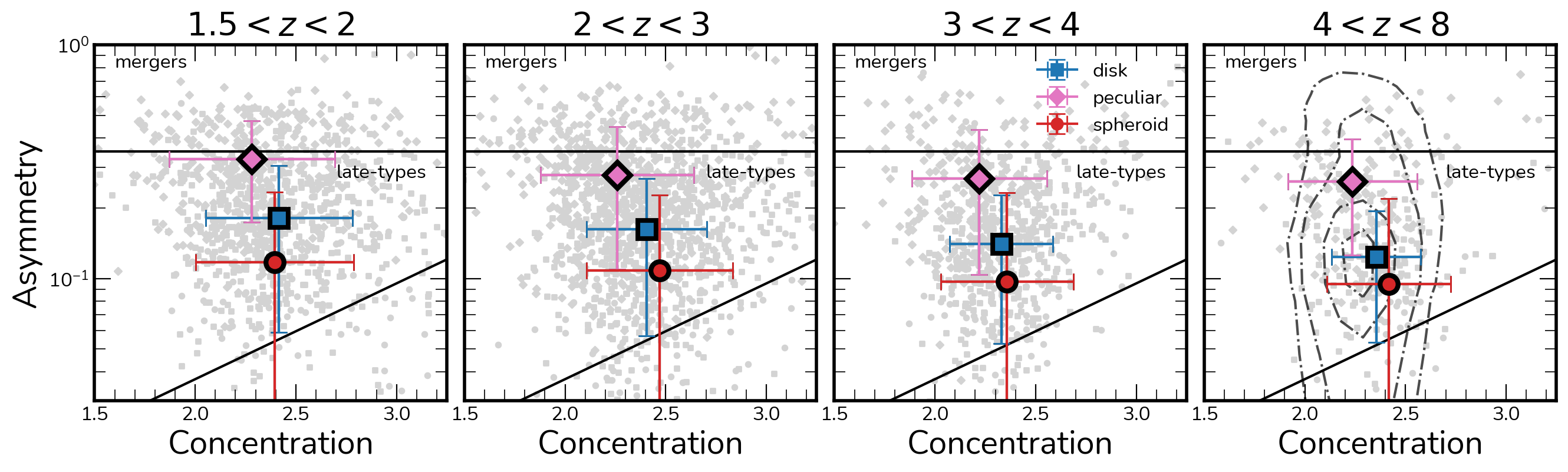}
    \caption{Concentration vs. Asymmetry diagrams. The evolution of the Concentration ($C$) and Asymmetry ($A$) in 4 different redshift bins, one for each panel. Peculiars are shown as pink diamonds, disks as blue squares and spheroids as red circles. The highest redshift panel shows contour lines based on the distribution of C-A measurements for BlueTides \citep{Madeline2022} galaxy mocks at $z\geq7$ as discussed in \S~\ref{sec:bluetides}.  The solid lines denote the merger selection threshold on the top, and the late type / early type separation on the lower diagonal line. Peculiars display high asymmetries when compared to other types. Disks display late type-like morphology, while spheroids are regular at lower redshifts, located on the bottom right of the plots, but move towards the center with increasing asymmetry and decreasing concentration with redshift. Galaxies overall get less concentrated and more asymmetric with increasing redshift. However, at high redshifts, sources display higher concentrations and asymmetries when compared to simulations.}
    \label{fig:CAS}
\end{figure*}

Figure~\ref{fig:CAS} shows the concentration ($C$) and asymmetry ($A$) plane based on \textsc{Morfometryka} measurements for 4 redshifts bins. The mean values alongside the distribution's $15\%$ and $85\%$ percentiles for disks, spheroids and peculiars are plotted as blue squares, red circles and pink diamonds, respectively. Each class has its distributions positioned within the expected regions for high redshift galaxies, with peculiars occupying the top of the diagram, the disks the central region and the spheroids around the lower right, including a high overlap. The positions of each of the three classes remain fairly stable over all redshifts, but the spheroids have higher asymmetries and lower concentrations overall at higher redshift, with larger overlaps with the disks. Also displayed with solid lines is the merger criterion based on asymmetry, as
\begin{equation}
    A > 0.35,
\end{equation}
and the diagonal late-type vs.~intermediate types boundary based on \cite{Bershady2000}.  We also explore the G-$M_{20}$ plane \citep{Lotz2004, Lotz2008} but we do not find any clear separation between the types, apart from the distinction between sources having close companions or being isolated, similar to what is reported in \cite{Rose2022}.

For the highest redshift bin ($4 < z < 8$), we also show the contours for the distribution of the measurements for 50,000 galaxy mocks from the BlueTides simulations \citep{Madeline2022}. The measurements from the real observations are in good agreement with the measurements for BlueTides mocks. We note that the effective spatial resolution of the BlueTides simulation is $1.5/h$ ckpc, which corresponds to 0.269 pkpc, or $0\farcs05$ at $z=7$. This resolution may have an effect on the resulting galaxy morphologies, particularly the inner regions and thus their concentration, but no clear effect is seen here, different to what is found for high-$z$ IllustrisTNG mocks \citep{Whitney2021}.

The spirality index -- the standard deviation of the galaxy polar  image $(r,\phi)$ gradient map, designed to measure the amount  of non-radial structures in the galaxy -- has proven to be very effective in discriminating different classes in this sample. If the galaxy is smooth, its polar image will consist of a single horizontal strip, which will imply a low value for $\sigma_\psi$. On the other hand, although we cannot resolve spiral arms in most cases, if the galaxy contains peripheral structures or  companions, the polar image will be irregular  with a corresponding high $\sigma_\psi$. We point the reader to the \textsc{Morfometryka} \citep{ferrari2015} paper for a full description of the $\sigma_\psi$ calculation. 

\begin{figure}
    \centering
    \includegraphics[width=0.40\textwidth]{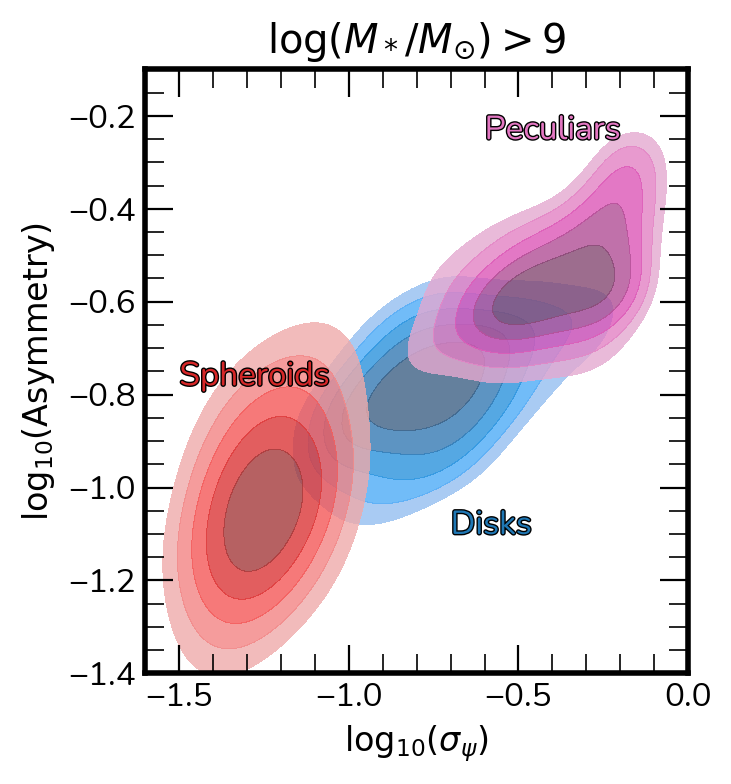}
    \caption{log Asymmetry ($A$) vs. log Spirality ($\sigma_{\psi}$) 2D distributions for each class. Kernel density estimation distributions for the top $50\%$ of each morphological class in 5 bins of 10\% fractions of the distribution. The asymmetry and $\sigma_\psi$ correlates strongly, but are independent measurements as each classification distribution has a different slope. Spheroids show high diversity in $A$ and low diversity in $\sigma_\psi$, while the contrary is true for disks. These two measurements form a parameter space capable of separating the classes in this sample relatively well compared to C-A.}
    \label{fig:spiralityA}
\end{figure}

\begin{figure*}
    \centering
    \includegraphics[width=0.7\textwidth]{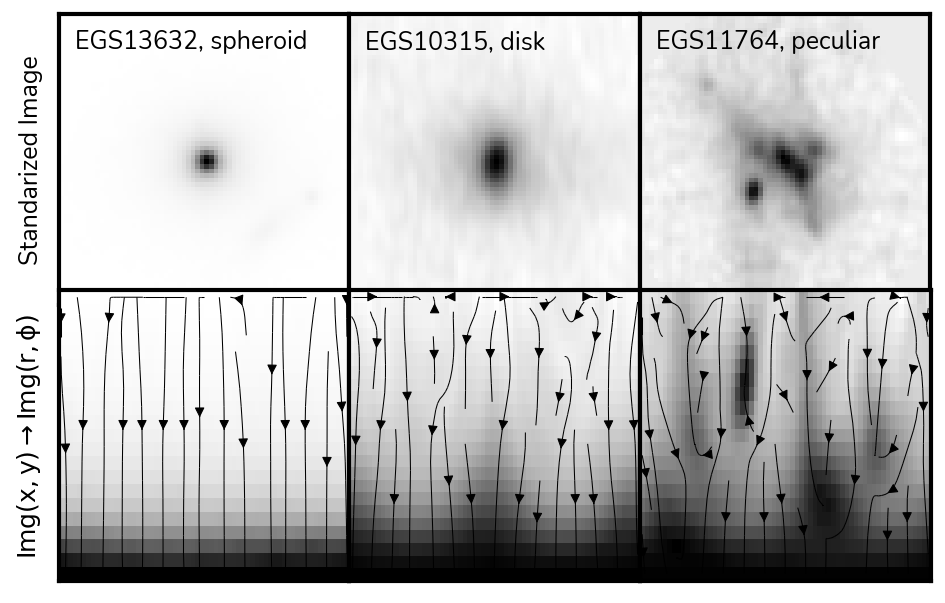}
    \caption{Three examples of the spirality $\sigma_\psi$ measurement. Top row shows the standarized images of the sources ($q=1$, $PA=0$) while the bottom row displays the polar coordinates transformation $(r, \phi)$ of the above image. Black lines display the gradient field of the image. The $\sigma_\psi$ measurement is based on the standard deviation of these field lines.}
    \label{fig:ex_spirality}
\end{figure*}

In Figure~\ref{fig:spiralityA} we show that combining $\sigma_\psi$ with the asymmetry ($A$) warrants a reasonable quantitative separation of the overall classes, as the center of each class distribution is well separated in this plane, unlike in the C-A or G-M20 diagram. As an example as to what is captured by the $\sigma_\psi$ measurement, we show three galaxies in Figure~\ref{fig:ex_spirality}, one for each class, with their respective polar coordinate image and the gradient lines that are used to compute $\sigma_\psi$. The distinction between spheroids and disks is subtle, but it is very powerful when large non-radial structure is present in the outskirts of the source.

\begin{figure}
    \centering
    \includegraphics[width=0.45\textwidth]{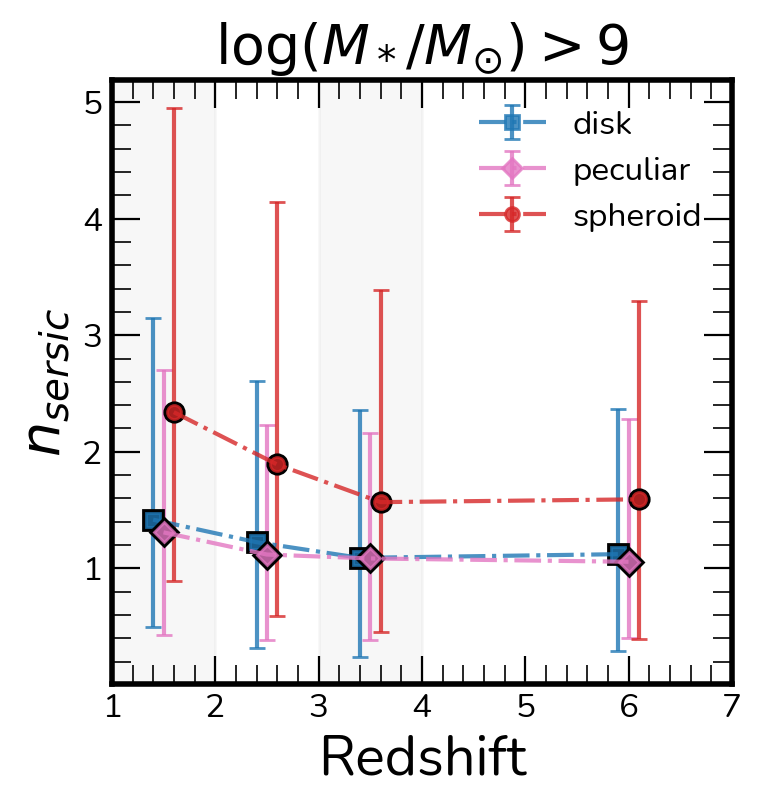}
    \caption{Sérsic index redshift evolution for each morphology class. Displayed as blue squares, red circles and pink 
diamonds are the means for disks, spheroids and peculiars, respectively. Error bars define the $15\%$ and $85\%$ percentile of the distributions.}
    \label{fig:sersic}
\end{figure}

Finally we explore the evolution of the Sérsic profiles \citep{Sersic1963} through the redshift evolution of the Sérsic index for galaxies with $M_* > 10^9 M_\odot$. In Figure~\ref{fig:sersic} we report mean values together with the $15\%$ and $85\%$ percentile limits as error bars to represent the distributions at each redshift bin for each class. Disks and Peculiars exhibit similar Sérsic profiles, with $n \approx 1.0$ at $z\sim6$ to $n \approx 1.3$ at $z \sim 1.5$. The spheroids, on the other hand, show higher Sérsic indicies at all redshifts, with $n\approx1.8$ at $z\sim6$ to $n\approx2.5$ at $z\sim1.5$. The distinction between the classes is clearer than what was reported in \cite{FERREIRA2022b} as the CANDELS overlap allows us to quickly select high mass galaxies only. The slopes for each class are also different, with the spheroids increasing in Sérsic index more rapidly. The similarities among disks and peculiars suggest that the majority of these disturbed and merging systems are still disk dominated.

\subsection{Evolution of the Hubble Sequence}\label{subsec:hubblesequence}

\begin{figure*}
    \centering
    \includegraphics[width=\textwidth]{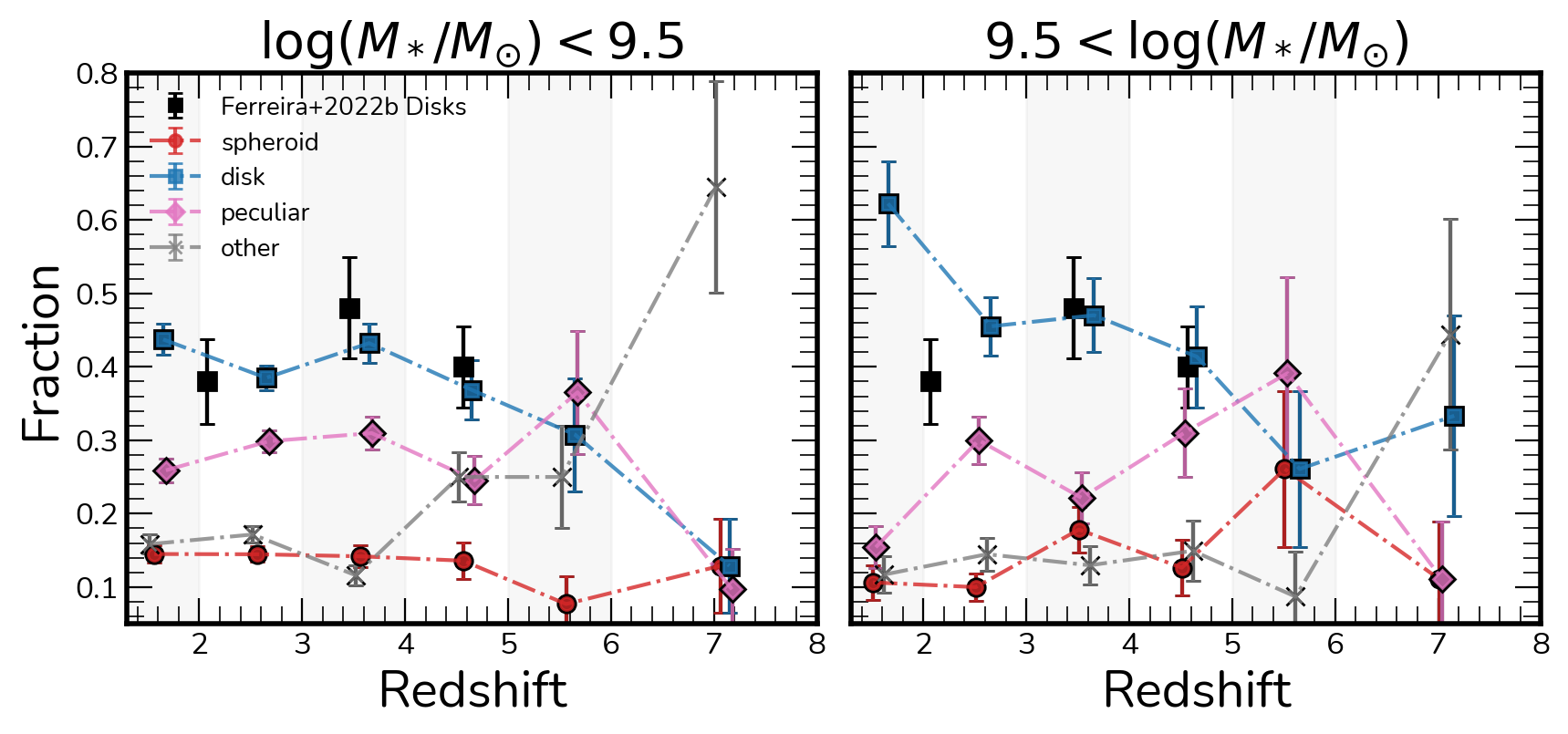}
    \caption{Morphology Fraction vs Redshift. The morphology fraction evolution with redshift for the main morphological classes of our classification framework (Fig.~\ref{fig:fluxogram}) in two mass bins, $\log(M_*/M_\odot) < 9.5$ ({\bf left}) and $\log(M_*/M_\odot) > 9.5$ ({\bf right}). Disks, spheroids, peculiars and other in blue squares, red circles, pink diamonds, and gray crosses, respectively. This \texttt{other} category aggregates the ambiguous, point source and unclassifiable sources.  The black squares show the Disk fractions reported in \citet{FERREIRA2022b}.}
    \label{fig:morpho_evo}
\end{figure*}

One principal goal of looking at galaxy morphology and structure is to establish when and how the Hubble Sequence \citep{Hubble1926} emerges in the context of the hierarchical assembly of the universe. 

Here we report the redshift evolution of morphological classes that encompass the three main categories of the Hubble Sequence from $1.5 < z < 8$, from when the Universe was only $\sim 0.6$ Gyrs old up to $\sim 4.2$ Gyrs. In Figure~\ref{fig:morpho_evo} we show this evolution in two mass bins, with the left panel displaying sources with $M_* \leq 10^{9.5} M\odot$ while the right panel shows this evolution for $M_* > 10^{9.5} M\odot$. Spheroids are displayed by red circles, peculiars as pink diamonds, disks as blue squares, and the other three possible categories as  \texttt{other} (point sources, ambiguous and unclassifiable) in gray crosses. The fraction of disks from \cite{FERREIRA2022b} are shown as black squares for comparison.

For low masses ($M_* \leq 10^{9.5} M\odot$) we do not find any systematic evolution with redshift, with all class fractions remaining fairly constant, with disks at $\sim 40\%$, peculiars at $\sim 25\%$ and spheroids at $\sim 15\%$. The only exception is at the highest redshift bin $6 < z < 8$, where all fractions go down to $\sim 10\%$ due to the sharp increase of faint and ambiguous sources, artifacts and unclassifiable cases, representing the limits on the depth of our dataset at the low mass range. The disk fractions between $2 < z < 5$ agrees well with what was previously reported in \cite{FERREIRA2022b}, but now with a 20 fold increase in sample size, as shown by the smaller error bars.

On the high mass cases ($M_* > 10^{9.5} M\odot$) we observe an evolution of the disk fraction from $\sim 55\%$ at $z\sim 2$ to $\sim 35\%$ at $z\sim7$ while the fractions of peculiars increases from $\sim 20\%$ to $\sim 30\%$ in the same redshift range, and spheroids increase from $\sim 10\%$ to $\sim 15\%$ between $1.5 < z < 6$.

We also note that there is a slight possible discrepancy between the number of high-mass disks between this work and our previous results in \cite{FERREIRA2022b}. This is most likely because the sample in \cite{FERREIRA2022b} includes galaxies of all mass, and so the sample is contaminated by low-mass galaxies which are less likely to be disks, as seen in our data as well.  However, when we divide our sample into masses and compare with \cite{FERREIRA2022b} we find very similar results.

\subsubsection{Spheroid Evolution}

Another remarkable aspect of the evolution of galaxy type fraction is that the spheroid fraction is roughly constant, even up to the highest redshifts, but lower than the level seen for local galaxies.  We generally do not see an increase in the number of elliptical galaxies at lower redshifts, at least down to $z \sim 1.5$.  We might have been expected to see more spheroids at lower-z if mergers are progressively transforming peculiars into ellipticals.  This indicates that elliptical galaxies at the mass ranges we probe must form morphologically relatively late in the history of the universe.  Despite this, there are clear indications that some ellipticals were morphologically present very early in the universe's history.  This may be a sign of different formation mechanisms at play for these spheroids at different epochs, with some forming through dissipative collapse, some from major mergers, and others through minor mergers.

These trends also suggest that the classic picture of morphology and structural evolution driven by merging might be only important for massive galaxies, where the low mass universe can be described broadly by a consistent Hubble Sequence in the range $1.5 < z < 6$. We discuss this in more detail in the section below.

\subsubsection{The role of mergers in early galaxy assembly}

Prior to \textit{JWST}, the scientific consensus obtained through both observations and simulations was that in the early Universe, galaxies grow and evolve through hierarchical mergers. In simulations, this is seen in galaxy merger trees \citep[e.g.,][]{Mo2010}, as well as in the fact that the merger rate increases with redshift \citep[e.g.,][]{Conselice2003,Rodriguez-Gomez2015, duncan2019, Ferreira2020}. Although observational studies of extremely high-redshift galaxies were previously limited, \cite{Mortlock2013} found an increase of the fraction of peculiar galaxies with redshift, supporting the idea that mergers were frequent.

However, the first JWST observations challenged this picture. \cite{FERREIRA2022b, NELSON2022, Jacobs2022, MORFEUS2022} all found a higher fraction of high-redshift disks and a lower fraction of peculiar galaxies than expected. The initial results, although based on small datasets, suggested that either mergers are less frequent that we thought, or that high-redshift disks tend to survive mergers and retain their disk morphology. 

In this paper, for the first time, we looked at a statistically large sample of high-redshift galaxies observed with \textit{JWST}, split into two different mass bins, to trace the evolution of high- and low-mass galaxies.

As stated in \ref{subsec:hubblesequence}, low-mass galaxies show a flat gradient in morphological type fractions. This suggests again that for low-mass galaxies, mergers do not drive structural evolution. Intuitively, this makes sense: since galaxies in this mass bin have assembled less mass, they are less likely to have experienced many mergers in their history. Therefore, the lack of obvious post-merger low-mass galaxies is not surprising.

In the high-mass bin, we see a steady decline of disk fraction at higher redshift, made up for by an increase of the peculiar fraction. This result supports the merger paradigm in principle and is similar to the pattern seen with HST \citep[e.g.,][]{Conselice2008}. Our results are in fact consistent with a significant fraction of high-mass galaxies undergoing mergers in the early Universe, going through a peculiar phase, and then forming stable disks.  Kinematic observations of these galaxies would go some way towards understanding if this picture is correct.

However, the fraction of high-mass disks is still significantly higher in our data than in \textit{HST}-based analyses \citep[e.g.,][]{Mortlock2013}. At $z=2$, $\sim$50\% of galaxies in our JWST observations have a disk morphology, compared to $\sim$10\% in \cite{Mortlock2013}. Therefore, our results still point at a tension between \textit{HST}- and \textit{JWST}-based morphology studies. Although our results are consistent with \textit{some} of the evolution between $z=1.5$ and $z=8$ being merger-driven, the fraction of $z>2$ disks is still too high.  We will investigate in future papers the detailed merger histories of these galaxies based on the JWST data. Ultimately, we want a self-consistent observational picture for how galaxy formation is occurring.

Overall, we see that peculiar structure is more common among high-mass than low-mass galaxies, pointing towards mass assembly via mergers; however, the fraction of disk galaxies in the high mass bin is still higher than what is observed by HST. Some implications of this result could be that either 1) galaxies grow by mergers as well as by another process, e.g., gas accretion \citep{L'Huillier2012}, or 2) mergers between high-mass galaxies do not destroy disks efficiently, allowing some galaxies to retain their disk morphology in the long-term \citep{Hopkins2009, Sparre2017}.

\subsection{Star Formation and Stellar Mass Evolution}\label{subsec:starformationevo}

\begin{figure*}
    \centering \includegraphics[width=\textwidth]{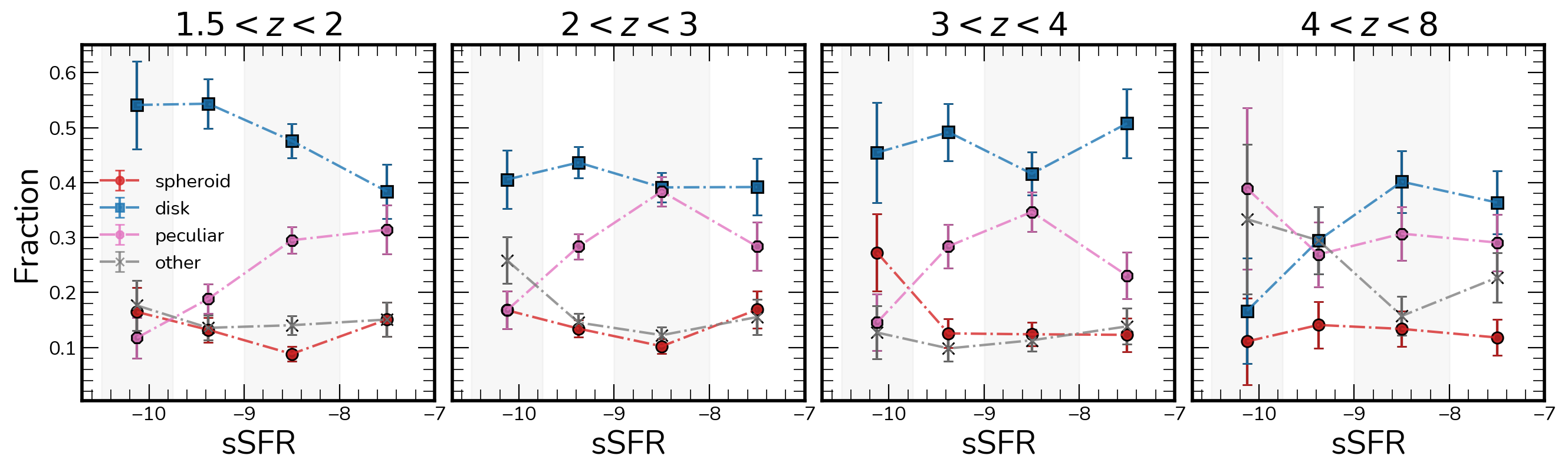}
    \caption{Morphology fractions vs average specific star formation. Disks, spheroids, peculiars and the other class are plotted in blue squares, red circles, pink diamonds, and gray crosses, respectively. Four redshift bins are shown and the \texttt{other} category aggregates the ambiguous, point source and unclassifiable sources. For all redshift bins and sSFR bins the disk galaxies dominate, with the exception of the highest redshift bin where disks and peculiar present similar contributions. }
    \label{fig:ssfr_evo}
\end{figure*}

One important goal of tracking morphologies across cosmic time is determining if different morphologies contribute to the star formation and the stellar mass budget of the universe differently. In the previous section we explored the fractional evolution with redshifts, concluding that disk galaxies dominate the overall fraction of morphologies at $1.5 < z < 6$, and possibly higher redshifts.  In Figure~\ref{fig:ssfr_evo} we show the class fractions in bins of specific star formation rates (sSFR) divided in four redshift bins. The overall fraction of galaxy types for each redshift panel can be seen in Fig.~\ref{fig:morpho_evo}, while each bin in Fig.~\ref{fig:ssfr_evo} shows these fractions for a given sSFR bin. Once more, the disks dominate the overall contribution in sSFR, with spheroids showing similar fraction to the overall fraction of spheroids in the sample. However, we see that at high sSFR bins, the contribution of peculiar galaxies increases, to become as important as disk galaxies despite their lower overall fractions in the sample. The same trend is shown for all redshift bins, as the fraction of peculiars increase with increasing sSFR. For the highest redshift bin ($4 < z < 8$), peculiars display roughly the same contribution as disks. Spheroid fractions are slightly higher at lower sSFR. This suggests that peculiar galaxies are important sites of star formation at all times in the Universe, and especially at higher redshifts.

\begin{figure}
    \centering
    \includegraphics[width=0.45\textwidth]{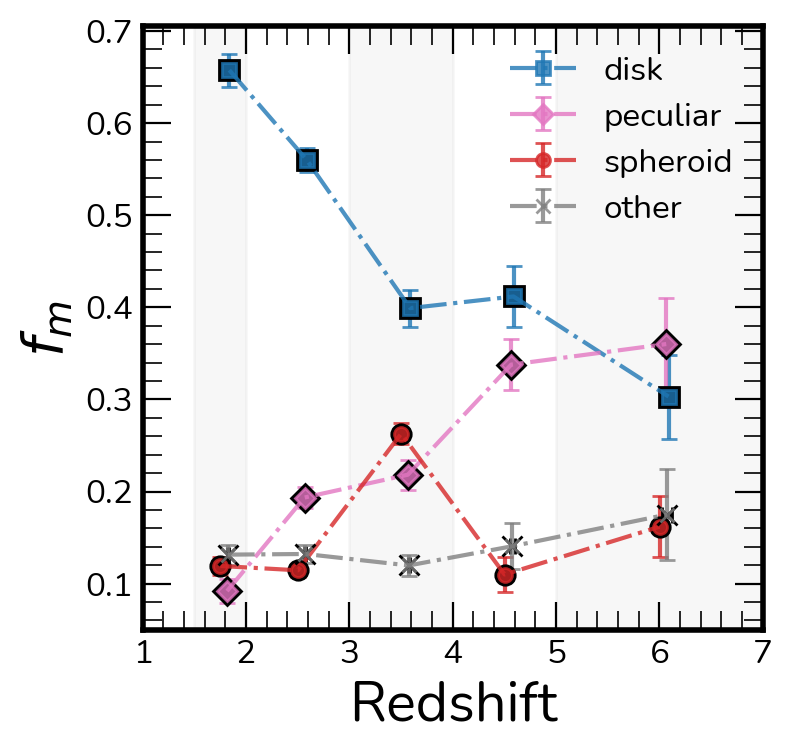}
    \caption{Fraction of the total stellar mass $(f_m)$ in each morphology subsample vs redshift.  This represents the relative amount each class contributes to the total stellar mass in that redshift. Disks, spheroids, peculiars, and the "other" class are plotted in blue squares, red circles, pink diamonds, and gray crosses, respectively. The \texttt{other} category aggregates the ambiguous, point source and unclassifiable sources. The contribution from disks shows a trade off with respect to peculiars and spheroids at higher redshifts. No stellar mass cut is applied, it includes all sample.}
    \label{fig:sf_evo}
\end{figure}

In Figure~\ref{fig:sf_evo} we show the contribution of each morphological class to the total stellar mass in each redshift bin. This shows that most of the mass in the very early universe was located in peculiar galaxies, while a clear trend with redshift is evident for disk galaxies, such that for $z<3$ most of the mass of the sample is distributed among disk galaxies. More massive overall individually, the spheroid galaxies hold just a small fraction of the total stellar mass in this sample ($f_m \sim 10\%$) as it is greatly outnumbered by the amount of mass in galaxies with disk morphologies. This is in contrast to what is found in the Hubble Deep Field \citep{Conselice2005} due to the very different morphology fractions reported.  It thus appears that for the bulk of the history of the universe, most stellar mass in the universe has been occupied in galaxies with disk like morphologies.

\begin{figure}
    \centering
    \includegraphics[width=0.45\textwidth]{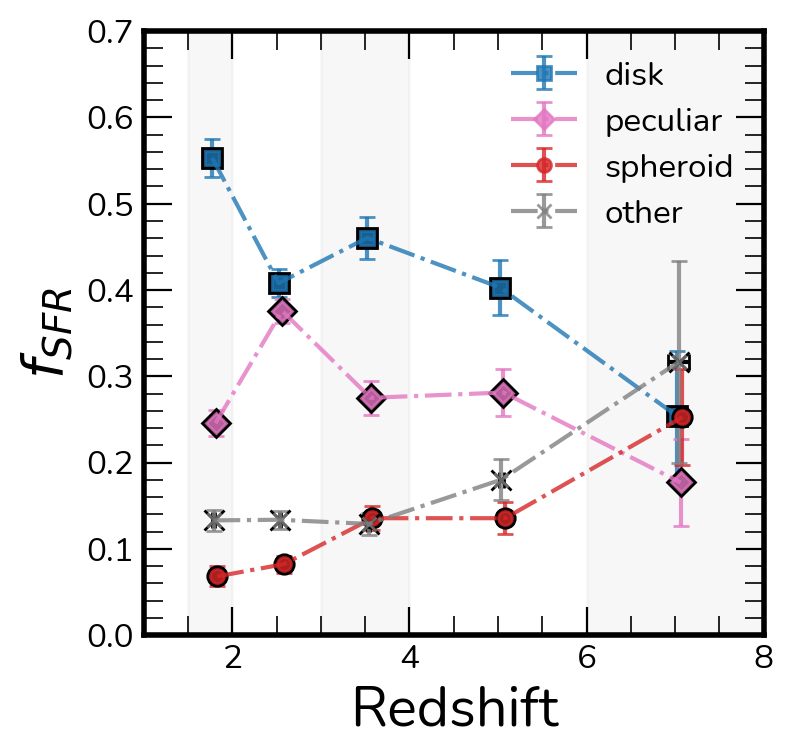}
    \caption{Fraction of SFR $(f_{SFR})$ in each morphology subsample vs redshift. 
    We show the contribution to total SFR of each redshift bin from each morphological class down to our mass limit. This represents how much each galaxy class contributes to the total SFR in that redshift. Disks, spheroids, peculiars, and the 'other' types are plotted in blue squares, red circles, pink diamonds, and gray crosses, respectively. No stellar mass cut is applied, it includes all sample.}
    \label{fig:sfr_evo}
\end{figure}

We plot in Figure~\ref{fig:sfr_evo} the fraction of star formation which exists in different galaxy types. As can be seen, the disk galaxies dominate the star formation rate in the universe up to at least $z \sim 6$.  What this implies is that, within the mass selection we use, the most likely galaxy type in which stars form are in disk galaxies.  However, this is still less than 50\% of the star formation until $z < 2$.  As can be seen, a significant amount of star formation also occurs in peculiar galaxies, which are probably mergers, but this does not dominate the method in which stars are formed.

\subsection{Predictions from Simulations}\label{subsec:simpreds}
We now look at numerical simulation results for morphological evolution over a similar redshift range. Simulations that resolve galaxies self-consistently typically model mass elements either on a grid or as particles. Particle--based decomposition methods \citep[e.g.][]{Abadi2003,Crain2010,Thob2019,Irodotou2021,Zana2022} have been extensively used in order to split galaxies into different morphological classes and facilitate a comparison between observed and simulated galactic properties \citep{Tissera2012,Pillepich2015,Irodotou2019,Monachesi2019,Trayford2019,RodriguezGomez2022}. However, the true morphology of a system may not always be accurately captured, as particle--based methods can be sensitive to small perturbations in the distribution of particles, which become progressively more significant at lower stellar masses as these galaxies are resolved with fewer particles.

\begin{figure}
    \centering
    \includegraphics[width=0.45\textwidth]{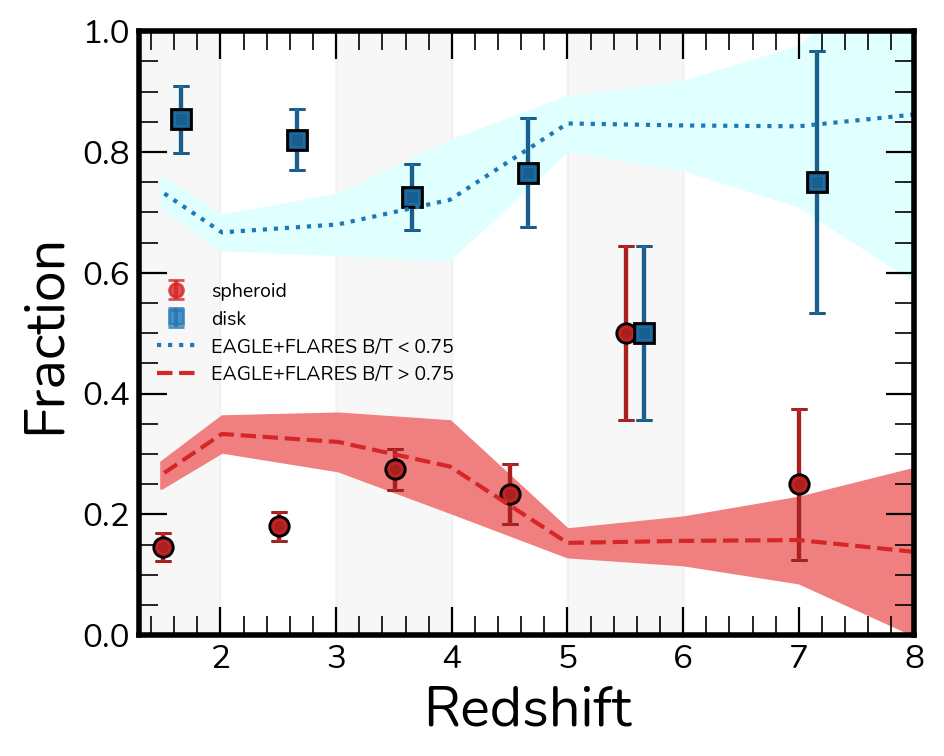}
    \caption{Morphology fractions compared to $B/T$ morphological type selection in EAGLE and FLARES for massive galaxies ($M_* \geq 10^{10} M_\odot$). The relative fraction of disks and spheroids are shown as blue squares and red circles, respectively. The blue dotted line shows mean values for galaxies with $B/T < 0.75$ in EAGLE and FLARES. Dashed line shows galaxies with $B/T > 0.75$. Shaded regions represents $\pm$2 sigma for each distribution. The visual classification fractions only account for disks and spheroids to allow a more direct comparison with the two thresholds in $B/T$ in the simulations. }
    \label{fig:fracsims}
\end{figure}

In this work, to ensure that both galaxies and their components are sufficiently resolved, and thus a particle--based decomposition is applicable, we use central and satellite galaxies from the \textsc{Eagle} \citep[][for $1.5<z<4$]{Schaye2015,Crain2015} and \textsc{Flares} \citep[][for $5<z<8$]{Lovell2021} simulations with stellar masses $\log(M_*/M_\odot) > 10$ (i.e.~even for a galaxy with $B/T \sim$ 0.2 the bulge is resolved with more than a thousand particles with mass of a few $\times\ 10^6$ \solm each). We use the method developed in \cite{Irodotou2021} to decompose galaxies by firstly creating a Mollweide projection of the angular momentum map of each galaxy’s stellar particles. Then, stellar particles are assigned to a disc or spheroid component based on their angular separation from the densest grid cell. This allows us to calculate bulge-to-total mass ratios ($B/T$) and use these to split galaxies into two morphological classes: (i) spheroids with $B/T>$ 0.75, (ii) spirals with $B/T<$ 0.75. The aforementioned $B/T$ limits were calibrated at $z\sim0$ in order for the \textsc{Eagle} galaxies to match the morphological classes in the \cite{Conselice2006} sample. In Figure~\ref{fig:fracsims} we show the comparison of these fractions with the relative fraction of disks and spheroids from the visual classifications for high mass galaxies with $M_* \geq 10^{10} M_\odot$, ignoring the peculiars, as we do not have a direct way to classify peculiars in the simulation dataset. The trends between simulations and visual classifications agree for $z>3$, with the exception of a single anomalous visual classification redshift bin showing similar fractions of disks and spheroids in $5 < z < 6$. Fractions for $z<3$ overall disagree, with an excess of $10\%$ of disks in the visual classifications. It is worth noting, however, in this redshift range (i.e. $1.5<z<3$), \citep{1701.04407} showed that the fraction of dry major mergers in the \textsc{Eagle} volume increases. Since this type of mergers can efficiently reduce the angular momentum of the remnant, this will translate in a negative correlation between our $B/T$ values and redshift, as also seen in e.g. Fig. 4 of \citep{1711.00030} for $M_* \geq 10^{10.5} M_\odot$.

\cite{Park2022} reports similar findings for the Horizon simulations, with the $z\geq5$ fraction of disks dominating at around $75\%$ by using quantitative morphology indicators as a proxy for morphology.

\subsection{HST vs JWST}\label{subsec:hstvjwst}

\begin{figure*}
    \centering
    \includegraphics[width=0.80\textwidth]{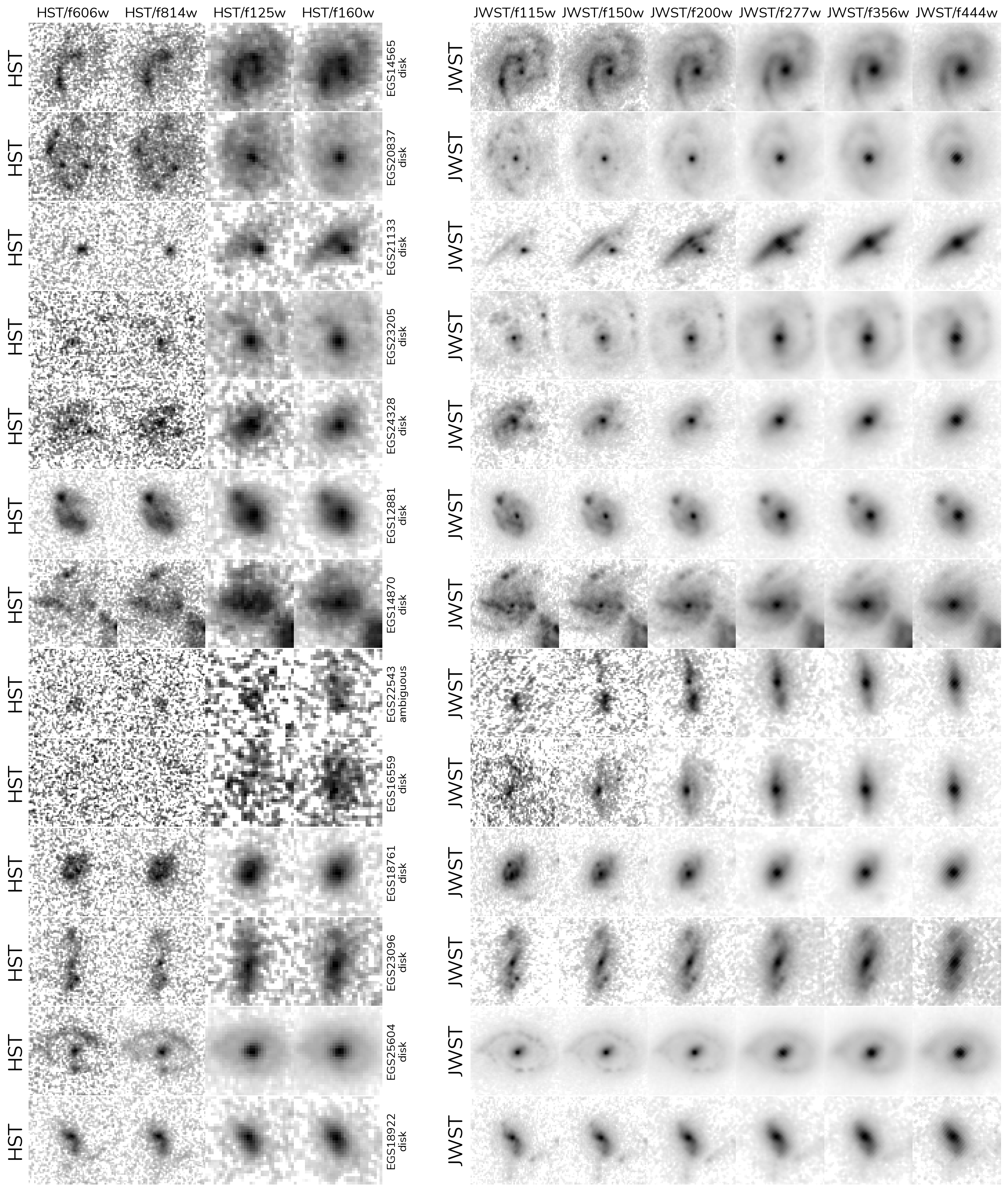}
    \caption{A HST vs.~JWST comparison. We show 13 galaxies in our sample that have observations in the four main CANDELS filters (left panel) and SW and LW filters in JWST (right panel). Faint features in CANDELS are generally very clear in JWST. In some cases only the central core of the galaxy is visible with the HST imaging. }
    \label{fig:hstvjwst}
\end{figure*}

As discussed in Sec.~\ref{subsec:hubblesequence}, there is a stark difference between morphological classifications derived with HST and JWST observations. The new JWST data is challenging our understanding of galaxy evolution and structure formation in the early Universe. Here, at the end of this paper, we discuss some differences between HST and JWST observations and provide an analysis of what contributes to this discrepancy.  This is meant to facilitate discussion about why our results are so different from previous work and to point the way towards understanding how to carry out future JWST galaxy structure work.

A comparison for a select number of galaxies in our sample between the NIRCam stamps and the HST ACS and WFC3 stamps is shown in Figure~\ref{fig:hstvjwst}. Many galaxies show very clear structures in NIRCam but ambiguous morphologies in HST. In some cases, such as EGS23205, only the central component is clearly seem in HST, while a disk, spiral arms and a bar pops up in the longer wavelength bands. In a few cases, such as EGS22543, the source is barely detected in the WFC3 and SW NIRCam images, while a clear disk is visible for the LW NIRCam stamps.

A detailed comparison with the CANDELS classifications \citep{KartaltepeCANDELS} is beyond the scope of this paper as our classifications do not align perfectly with the scheme defined in \cite{KartaltepeCANDELS}. However, we briefly discuss the modes for which the discrepancies between classifications based on HST imaging and NIRCam can be explained. First, many of the fine structure such as bars and spirals, are hard to resolve at high redshifts due to the WFC3 pixel scale, and can be mistaken by merging signatures or disturbances, such as the case of EGS14565. It shows clear spiral structure in JWST but not as clear in HST, which could be mistaken as a merger in HST. Secondly, we find the wavelength coverage to be critical, as many of the galaxies at high redshift in HST are probed in the blue side of the optical, and are prone to absorption from dust, giving rise to asymmetric looking structures. Moreover, the bluer bands probe the youngest stars, which have more irregular spatial distributions tracing sites of ongoing star formation but not the underlying mass distribution.  Galaxies such as EGS522543 and EGS16559 are good examples of this. This is also expected to be an issue for galaxies at $z>7$ in NIRCam images, as we start having the same issue as WFC3 for $2 < z < 3$.

Future morphology classification studies, with more detailed descriptions and covering larger datasets, such as the complete cycle of CEERS, PRIMER, COSMOS-WEB, JADES and NGDEEP, together with the scope of citizen science projects like Galaxy Zoo \citep{Lintott2011}, will enable a detailed discussion on the main differences between HST and JWST morphology in the overlap region $2 < z < 3$.

\section{Summary and Conclusions}\label{sec:summary}

We present results on the rest-frame optical morphologies and the structural evolution of JWST observed galaxies at $z=1.5$ to 8 in a statistically significant sample of 4265 galaxies for the first time, using both visual classifications and quantitative morphology. We focus on galaxies observed by the CEERS program that overlap with the CANDELS fields, enabling us to use robust measurements of redshifts, stellar masses and star formation rates available in CANDELS.

Our major findings are:

I. Distant galaxies at $z > 1.5$ display surprisingly regular disk morphologies at early times, such that for galaxies with $M_* < 10^{9.5} M_\odot$, the fraction of disks/spheroids/peculiars appear to to be roughly constant at $1.5 < z < 6$, showing that the Hubble Sequence was already in place as early as one billion years after the Big Bang. 

II. For galaxies with higher masses $M_* > 10^{9.5} M_\odot$, tremendous evolution is observed in the fraction of disks and peculiars, suggesting that the role of mergers might be more important to the massive cases.

III. Non-parametric morphology measurements agree well with visual classifications. However, a large overlap exists between classes in the usual CAS, etc. planes. We find that the spirality index ($\sigma_\psi$) when combined with the asymmetry ($A$) makes a powerful diagnostic to separate disks/spheroids/peculiars. 

IV. Comparisons with B/T studies from EAGLE and FLARES show that quantitative structures at high redshifts agree well with simulations, and are not unexpected from a theoretical standpoint, even if discrepant with previous morphological studies with HST. 

V. Galaxies with disk morphologies dominate both the low sSFR and high sSFR populations, fairly outnumbering spheroids. However, the peculiar contribution to the sSFR budget increases with increasing sSFR and redshift, such that at the highest redshifts, the majority of the highly star forming galaxy population has disturbed/peculiar morphologies.

VI. The contribution to the total stellar mass of galaxies at high redshift is dominated by peculiar galaxies, while most of the stellar mass in the Universe at $z<3$ is located in disk galaxies.  We also find that disk galaxies dominate the star formation rate at $z < 6$, suggesting that most stars in the universe were formed in a galaxy with a disk morphology.
VII. We report clear examples of galaxies whose morphologies are hidden in HST imaging, but become clear and unambiguous in the NIRCam observations. Spirals and bars are better resolved and clear in the LW NIRCam filters.

In addition to the morphology study presented in this paper, we release the first version of our catalog of aggregated classifications to the community. Our goal is that this large sample of visually classified galaxies will serve as a base for early studies on morphology and structure, and will help the community develop methods and tools to tackle scheduled larger area observations such as COSMOS-WEB while data releases from large citizen science classifications projects, such as GalaxyZoo, are not available. These classifications can be used, for example, as an early training dataset for deep learning methods, or as a transfer learning sample for already established models. 

\acknowledgements

We thank Anthony Holloway, Sotirios Sanidas and Phil Perry for critical and timely help with computer infrastructure that made this work possible. We acknowledge support from the ERC Advanced Investigator Grant EPOCHS (788113), as well as a studentship from STFC. LF acknowledges financial support from Coordenação de Aperfeiçoamento de Pessoal de Nível Superior - Brazil (CAPES) in the form of a PhD studentship. DI acknowledges support by the European Research Council via ERC Consolidator Grant KETJU (no. 818930). CCL acknowledges support from the Royal Society under grant RGF/EA/181016. CT acknowledges funding from the Science and Technology Facilities Council (STFC). This work is based on observations made with the NASA/ESA \textit{Hubble Space Telescope} (HST) and NASA/ESA/CSA \textit{James Webb Space Telescope} (JWST) obtained from the \texttt{Mikulski Archive for Space Telescopes} (\texttt{MAST}) at the \textit{Space Telescope Science Institute} (STScI), which is operated by the Association of Universities for Research in Astronomy, Inc., under NASA contract NAS 5-03127 for JWST, and NAS 5–26555 for HST.

This research made use of the following Python libraries: \textsc{Astropy} \citep{astropy2022}; \textsc{Morfometryka} \citep{ferrari2015}; \textsc{Pandas} \citep{pandas}; \textsc{Matplotlib} \citep{Hunter:2007}; \textsc{photutils} \citep{larry_bradley_2020_4044744}

\bibliography{corrected_refs}{}

\begin{thebibliography}{}
\expandafter\ifx\csname natexlab\endcsname\relax\def\natexlab#1{#1}\fi
\providecommand{\url}[1]{\href{#1}{#1}}
\providecommand{\dodoi}[1]{doi:~\href{http://doi.org/#1}{\nolinkurl{#1}}}
\providecommand{\doeprint}[1]{\href{http://ascl.net/#1}{\nolinkurl{http://ascl.net/#1}}}
\providecommand{\doarXiv}[1]{\href{https://arxiv.org/abs/#1}{\nolinkurl{https://arxiv.org/abs/#1}}}

\bibitem[{{Abadi} {et~al.}(2003){Abadi}, {Navarro}, {Steinmetz}, \&
  {Eke}}]{Abadi2003}
{Abadi}, M.~G., {Navarro}, J.~F., {Steinmetz}, M., \& {Eke}, V.~R. 2003, \apj,
  591, 499, \dodoi{10.1086/375512}

\bibitem[{Abraham {et~al.}(1994)Abraham, Valdes, Yee, \& van~den
  Bergh}]{Abraham1994}
Abraham, R.~G., Valdes, F., Yee, H. K.~C., \& van~den Bergh, S. 1994, The
  Astrophysical Journal, 432, 75, \dodoi{10.1086/174550}

\bibitem[{{Adams} {et~al.}(2022){Adams}, {Conselice}, {Ferreira}, {Austin},
  {Trussler}, {Juod{\v{z}}balis}, {Wilkins}, {Caruana}, {Dayal}, {Verma}, \&
  {Vijayan}}]{adams2022}
{Adams}, N.~J., {Conselice}, C.~J., {Ferreira}, L., {et~al.} 2022, arXiv
  e-prints, arXiv:2207.11217.
\newblock \doarXiv{2207.11217}

\bibitem[{{Ashby} {et~al.}(2015){Ashby}, {Willner}, {Fazio}, {Dunlop}, {Egami},
  {Faber}, {Ferguson}, {Grogin}, {Hora}, {Huang}, {Koekemoer}, {Labb{\'e}}, \&
  {Wang}}]{Ashby2015}
{Ashby}, M.~L.~N., {Willner}, S.~P., {Fazio}, G.~G., {et~al.} 2015, \apjs, 218,
  33, \dodoi{10.1088/0067-0049/218/2/33}

\bibitem[{{Astropy Collaboration} {et~al.}(2022){Astropy Collaboration},
  {Price-Whelan}, {Lim}, {Earl}, {Starkman}, {Bradley}, {Shupe}, {Patil},
  {Corrales}, {Brasseur}, {N{\"o}the}, {Donath}, {Tollerud}, {Morris},
  {Ginsburg}, {Vaher}, {Weaver}, {Tocknell}, {Jamieson}, {van Kerkwijk},
  {Robitaille}, {Merry}, {Bachetti}, {G{\"u}nther}, {Aldcroft},
  {Alvarado-Montes}, {Archibald}, {B{\'o}di}, {Bapat}, {Barentsen},
  {Baz{\'a}n}, {Biswas}, {Boquien}, {Burke}, {Cara}, {Cara}, {Conroy},
  {Conseil}, {Craig}, {Cross}, {Cruz}, {D'Eugenio}, {Dencheva}, {Devillepoix},
  {Dietrich}, {Eigenbrot}, {Erben}, {Ferreira}, {Foreman-Mackey}, {Fox},
  {Freij}, {Garg}, {Geda}, {Glattly}, {Gondhalekar}, {Gordon}, {Grant},
  {Greenfield}, {Groener}, {Guest}, {Gurovich}, {Handberg}, {Hart},
  {Hatfield-Dodds}, {Homeier}, {Hosseinzadeh}, {Jenness}, {Jones}, {Joseph},
  {Kalmbach}, {Karamehmetoglu}, {Ka{\l}uszy{\'n}ski}, {Kelley}, {Kern},
  {Kerzendorf}, {Koch}, {Kulumani}, {Lee}, {Ly}, {Ma}, {MacBride}, {Maljaars},
  {Muna}, {Murphy}, {Norman}, {O'Steen}, {Oman}, {Pacifici}, {Pascual},
  {Pascual-Granado}, {Patil}, {Perren}, {Pickering}, {Rastogi}, {Roulston},
  {Ryan}, {Rykoff}, {Sabater}, {Sakurikar}, {Salgado}, {Sanghi}, {Saunders},
  {Savchenko}, {Schwardt}, {Seifert-Eckert}, {Shih}, {Jain}, {Shukla}, {Sick},
  {Simpson}, {Singanamalla}, {Singer}, {Singhal}, {Sinha}, {Sip{\H{o}}cz},
  {Spitler}, {Stansby}, {Streicher}, {{\v{S}}umak}, {Swinbank}, {Taranu},
  {Tewary}, {Tremblay}, {Val-Borro}, {Van Kooten}, {Vasovi{\'c}}, {Verma}, {de
  Miranda Cardoso}, {Williams}, {Wilson}, {Winkel}, {Wood-Vasey}, {Xue},
  {Yoachim}, {Zhang}, {Zonca}, \& {Astropy Project Contributors}}]{astropy2022}
{Astropy Collaboration}, {Price-Whelan}, A.~M., {Lim}, P.~L., {et~al.} 2022,
  \apj, 935, 167, \dodoi{10.3847/1538-4357/ac7c74}

\bibitem[{Bershady {et~al.}(2000)Bershady, Jangren, \&
  Conselice}]{Bershady2000}
Bershady, M.~A., Jangren, A., \& Conselice, C.~J. 2000, The Astronomical
  Journal, 119, 2645, \dodoi{10.1086/301386}

\bibitem[{Bradley {et~al.}(2020)Bradley, Sip{\H o}cz, Robitaille, Tollerud,
  Vin{\'{\i}}cius, Deil, Barbary, Wilson, Busko, G{\"u}nther, Cara, Conseil,
  Bostroem, Droettboom, Bray, Bratholm, Lim, Barentsen, Craig, Pascual, Perren,
  Greco, Donath, de~Val-Borro, Kerzendorf, Bach, Weaver, D'Eugenio, Souchereau,
  \& Ferreira}]{larry_bradley_2020_4044744}
Bradley, L., Sip{\H o}cz, B., Robitaille, T., {et~al.} 2020, astropy/photutils:
  1.0.0, 1.0.0,  Zenodo, \dodoi{10.5281/zenodo.4044744}

\bibitem[{Buitrago {et~al.}(2008)Buitrago, Trujillo, Conselice, Bouwens,
  Dickinson, \& Yan}]{Buitrago2008}
Buitrago, F., Trujillo, I., Conselice, C.~J., {et~al.} 2008, The Astrophysical
  Journal, 687, L61, \dodoi{10.1086/592836}

\bibitem[{{Buitrago} {et~al.}(2013){Buitrago}, {Trujillo}, {Conselice}, \&
  {H{\"a}u{\ss}ler}}]{Buitrago2013}
{Buitrago}, F., {Trujillo}, I., {Conselice}, C.~J., \& {H{\"a}u{\ss}ler}, B.
  2013, \mnras, 428, 1460, \dodoi{10.1093/mnras/sts124}

\bibitem[{{Clauwens} {et~al.}(2018){Clauwens}, {Schaye}, {Franx}, \&
  {Bower}}]{1711.00030}
{Clauwens}, B., {Schaye}, J., {Franx}, M., \& {Bower}, R.~G. 2018, \mnras, 478,
  3994, \dodoi{10.1093/mnras/sty1229}

\bibitem[{{Coe} {et~al.}(2015){Coe}, {Bradley}, \& {Zitrin}}]{Coe2015}
{Coe}, D., {Bradley}, L., \& {Zitrin}, A. 2015, \apj, 800, 84,
  \dodoi{10.1088/0004-637X/800/2/84}

\bibitem[{Conselice(2003)}]{Conselice2003a}
Conselice, C.~J. 2003, The Astrophysical Journal Supplement Series, 147, 1,
  \dodoi{10.1086/375001}

\bibitem[{Conselice(2006)}]{Conselice2006}
---. 2006, The Astrophysical Journal, 638, 686, \dodoi{10.1086/499067}

\bibitem[{Conselice(2014)}]{Conselice2014a}
---. 2014, Annual Review of Astronomy and Astrophysics, 52, 291,
  \dodoi{10.1146/annurev-astro-081913-040037}

\bibitem[{{Conselice} \& {Arnold}(2009)}]{conselice2009}
{Conselice}, C.~J., \& {Arnold}, J. 2009, \mnras, 397, 208,
  \dodoi{10.1111/j.1365-2966.2009.14959.x}

\bibitem[{Conselice {et~al.}(2003)Conselice, Bershady, Dickinson, \&
  Papovich}]{Conselice2003}
Conselice, C.~J., Bershady, M.~A., Dickinson, M., \& Papovich, C. 2003, The
  Astronomical Journal, 126, 1183, \dodoi{10.1086/377318}

\bibitem[{{Conselice} {et~al.}(2005){Conselice}, {Blackburne}, \&
  {Papovich}}]{Conselice2005}
{Conselice}, C.~J., {Blackburne}, J.~A., \& {Papovich}, C. 2005, \apj, 620,
  564, \dodoi{10.1086/426102}

\bibitem[{Conselice {et~al.}(2008)Conselice, Rajgor, \& Myers}]{Conselice2008}
Conselice, C.~J., Rajgor, S., \& Myers, R. 2008, Monthly Notices of the Royal
  Astronomical Society, 386, 909, \dodoi{10.1111/j.1365-2966.2008.13069.x}

\bibitem[{{Crain} {et~al.}(2010){Crain}, {McCarthy}, {Frenk}, {Theuns}, \&
  {Schaye}}]{Crain2010}
{Crain}, R.~A., {McCarthy}, I.~G., {Frenk}, C.~S., {Theuns}, T., \& {Schaye},
  J. 2010, \mnras, 407, 1403, \dodoi{10.1111/j.1365-2966.2010.16985.x}

\bibitem[{{Crain} {et~al.}(2015){Crain}, {Schaye}, {Bower}, {Furlong},
  {Schaller}, {Theuns}, {Dalla Vecchia}, {Frenk}, {McCarthy}, {Helly},
  {Jenkins}, {Rosas-Guevara}, {White}, \& {Trayford}}]{Crain2015}
{Crain}, R.~A., {Schaye}, J., {Bower}, R.~G., {et~al.} 2015, \mnras, 450, 1937,
  \dodoi{10.1093/mnras/stv725}

\bibitem[{{Delgado-Serrano} {et~al.}(2010){Delgado-Serrano}, {Hammer}, {Yang},
  {Puech}, {Flores}, \& {Rodrigues}}]{Delgado-Serrano2010}
{Delgado-Serrano}, R., {Hammer}, F., {Yang}, Y.~B., {et~al.} 2010, \aap, 509,
  A78, \dodoi{10.1051/0004-6361/200912704}

\bibitem[{{Driver} {et~al.}(1995){Driver}, {Windhorst}, \&
  {Griffiths}}]{Driver1995}
{Driver}, S.~P., {Windhorst}, R.~A., \& {Griffiths}, R.~E. 1995, \apj, 453, 48,
  \dodoi{10.1086/176369}

\bibitem[{Duncan {et~al.}(2014)Duncan, Conselice, Mortlock, Hartley, Guo,
  Ferguson, Dave, Lu, Ownsworth, Ashby, Dekel, Dickinson, Faber, Giavalisco,
  Grogin, Kocevski, Koekemoer, Somerville, \& White}]{Duncan2014}
Duncan, K., Conselice, C.~J., Mortlock, A., {et~al.} 2014, Monthly Notices of
  the Royal Astronomical Society, 444, 2960, \dodoi{10.1093/mnras/stu1622}

\bibitem[{{Duncan} {et~al.}(2019){Duncan}, {Conselice}, {Mundy}, {Bell},
  {Donley}, {Galametz}, {Guo}, {Grogin}, {Hathi}, {Kartaltepe}, {Kocevski},
  {Koekemoer}, {P{\'e}rez-Gonz{\'a}lez}, {Mantha}, {Snyder}, \&
  {Stefanon}}]{duncan2019}
{Duncan}, K., {Conselice}, C.~J., {Mundy}, C., {et~al.} 2019, \apj, 876, 110,
  \dodoi{10.3847/1538-4357/ab148a}

\bibitem[{Feng {et~al.}(2015)Feng, Di-Matteo, Croft, Bird, Battaglia, \&
  Wilkins}]{Feng2015}
Feng, Y., Di-Matteo, T., Croft, R.~A., {et~al.} 2015, MNRAS, 455, 2778,
  \dodoi{10.1093/mnras/stv2484}

\bibitem[{{Ferrari} {et~al.}(2015){Ferrari}, {de Carvalho}, \&
  {Trevisan}}]{ferrari2015}
{Ferrari}, F., {de Carvalho}, R.~R., \& {Trevisan}, M. 2015, \apj, 814, 55,
  \dodoi{10.1088/0004-637X/814/1/55}

\bibitem[{{Ferreira} {et~al.}(2020){Ferreira}, {Conselice}, {Duncan}, {Cheng},
  {Griffiths}, \& {Whitney}}]{Ferreira2020}
{Ferreira}, L., {Conselice}, C.~J., {Duncan}, K., {et~al.} 2020, \apj, 895,
  115, \dodoi{10.3847/1538-4357/ab8f9b}

\bibitem[{{Ferreira} \& {Ferrari}(2018)}]{Ferreira2018}
{Ferreira}, L., \& {Ferrari}, F. 2018, \mnras, 473, 2701,
  \dodoi{10.1093/mnras/stx2266}

\bibitem[{{Ferreira} {et~al.}(2022){Ferreira}, {Adams}, {Conselice},
  {Sazonova}, {Austin}, {Caruana}, {Ferrari}, {Verma}, {Trussler},
  {Broadhurst}, {Diego}, {Frye}, {Pascale}, {Wilkins}, {Windhorst}, \&
  {Zitrin}}]{FERREIRA2022b}
{Ferreira}, L., {Adams}, N., {Conselice}, C.~J., {et~al.} 2022, arXiv e-prints,
  arXiv:2207.09428.
\newblock \doarXiv{2207.09428}

\bibitem[{Grogin {et~al.}(2011)Grogin, Kocevski, Faber, Ferguson, Koekemoer,
  Riess, Acquaviva, Alexander, Almaini, Ashby, Barden, Bell, Bournaud, Brown,
  Caputi, Casertano, Cassata, Castellano, Challis, Chary, Cheung, Cirasuolo,
  Conselice, Cooray, Croton, Daddi, Dahlen, Dav{\'{e}}, de~Mello, Dekel,
  Dickinson, Dolch, Donley, Dunlop, Dutton, Elbaz, Fazio, Filippenko,
  Finkelstein, Fontana, Gardner, Garnavich, Gawiser, Giavalisco, Grazian, Guo,
  Hathi, H{\"{a}}ussler, Hopkins, Huang, Huang, Jha, Kartaltepe, Kirshner, Koo,
  Lai, Lee, Li, Lotz, Lucas, Madau, McCarthy, McGrath, McIntosh, McLure,
  Mobasher, Moustakas, Mozena, Nandra, Newman, Niemi, Noeske, Papovich,
  Pentericci, Pope, Primack, Rajan, Ravindranath, Reddy, Renzini, Rix, Robaina,
  Rodney, Rosario, Rosati, Salimbeni, Scarlata, Siana, Simard, Smidt,
  Somerville, Spinrad, Straughn, Strolger, Telford, Teplitz, Trump, van~der
  Wel, Villforth, Wechsler, Weiner, Wiklind, Wild, Wilson, Wuyts, Yan, \&
  Yun}]{Grogin2011}
Grogin, N.~A., Kocevski, D.~D., Faber, S.~M., {et~al.} 2011, The Astrophysical
  Journal Supplement Series, 197, 35, \dodoi{10.1088/0067-0049/197/2/35}

\bibitem[{{Hopkins} {et~al.}(2009){Hopkins}, {Cox}, {Younger}, \&
  {Hernquist}}]{Hopkins2009}
{Hopkins}, P.~F., {Cox}, T.~J., {Younger}, J.~D., \& {Hernquist}, L. 2009,
  \apj, 691, 1168, \dodoi{10.1088/0004-637X/691/2/1168}

\bibitem[{Hubble(1926)}]{Hubble1926}
Hubble, E.~P. 1926, The Astrophysical Journal, 64, 321, \dodoi{10.1086/143018}

\bibitem[{Hunter(2007)}]{Hunter:2007}
Hunter, J.~D. 2007, Computing in Science \& Engineering, 9, 90,
  \dodoi{10.1109/MCSE.2007.55}

\bibitem[{{Irodotou} \& {Thomas}(2021)}]{Irodotou2021}
{Irodotou}, D., \& {Thomas}, P.~A. 2021, \mnras, 501, 2182,
  \dodoi{10.1093/mnras/staa3804}

\bibitem[{{Irodotou} {et~al.}(2019){Irodotou}, {Thomas}, {Henriques},
  {Sargent}, \& {Hislop}}]{Irodotou2019}
{Irodotou}, D., {Thomas}, P.~A., {Henriques}, B.~M., {Sargent}, M.~T., \&
  {Hislop}, J.~M. 2019, \mnras, 489, 3609, \dodoi{10.1093/mnras/stz2365}

\bibitem[{{Jacobs} {et~al.}(2022){Jacobs}, {Glazebrook}, {Calabr{\`o}}, {Treu},
  {Nanayakkara}, {Jones}, {Merlin}, {Abraham}, {Stevens}, {Vulcani}, {Yang},
  {Bonchi}, {Bradac}, {Castellano}, {Fontana}, {Mason}, {Morishita}, {Paris},
  {Trenti}, {Marchesini}, {Wang}, \& {Santini}}]{Jacobs2022}
{Jacobs}, C., {Glazebrook}, K., {Calabr{\`o}}, A., {et~al.} 2022, arXiv
  e-prints, arXiv:2208.06516.
\newblock \doarXiv{2208.06516}

\bibitem[{{Kartaltepe} {et~al.}(2015){Kartaltepe}, {Mozena}, {Kocevski},
  {McIntosh}, \& {Lotz}}]{KartaltepeCANDELS}
{Kartaltepe}, J.~S., {Mozena}, M., {Kocevski}, D., {McIntosh}, D.~H., \&
  {Lotz}, J. 2015, \apjs, 221, 11, \dodoi{10.1088/0067-0049/221/1/11}

\bibitem[{{Koekemoer} {et~al.}(2011){Koekemoer}, {Faber}, {Ferguson}, {Grogin},
  {Kocevski}, {Koo}, {Lai}, {Lotz}, {Lucas}, {McGrath}, {Ogaz}, {Rajan},
  {Riess}, {Rodney}, {Strolger}, {Casertano}, {Castellano}, {Dahlen},
  {Dickinson}, {Dolch}, {Fontana}, {Giavalisco}, {Grazian}, {Guo}, {Hathi},
  {Huang}, {van der Wel}, {Yan}, {Acquaviva}, {Alexander}, {Almaini}, {Ashby},
  {Barden}, {Bell}, {Bournaud}, {Brown}, {Caputi}, {Cassata}, {Challis},
  {Chary}, {Cheung}, {Cirasuolo}, {Conselice}, {Roshan Cooray}, {Croton},
  {Daddi}, {Dav{\'e}}, {de Mello}, {de Ravel}, {Dekel}, {Donley}, {Dunlop},
  {Dutton}, {Elbaz}, {Fazio}, {Filippenko}, {Finkelstein}, {Frazer}, {Gardner},
  {Garnavich}, {Gawiser}, {Gruetzbauch}, {Hartley}, {H{\"a}ussler},
  {Herrington}, {Hopkins}, {Huang}, {Jha}, {Johnson}, {Kartaltepe},
  {Khostovan}, {Kirshner}, {Lani}, {Lee}, {Li}, {Madau}, {McCarthy},
  {McIntosh}, {McLure}, {McPartland}, {Mobasher}, {Moreira}, {Mortlock},
  {Moustakas}, {Mozena}, {Nandra}, {Newman}, {Nielsen}, {Niemi}, {Noeske},
  {Papovich}, {Pentericci}, {Pope}, {Primack}, {Ravindranath}, {Reddy},
  {Renzini}, {Rix}, {Robaina}, {Rosario}, {Rosati}, {Salimbeni}, {Scarlata},
  {Siana}, {Simard}, {Smidt}, {Snyder}, {Somerville}, {Spinrad}, {Straughn},
  {Telford}, {Teplitz}, {Trump}, {Vargas}, {Villforth}, {Wagner}, {Wandro},
  {Wechsler}, {Weiner}, {Wiklind}, {Wild}, {Wilson}, {Wuyts}, \&
  {Yun}}]{Koekemoer2011}
{Koekemoer}, A.~M., {Faber}, S.~M., {Ferguson}, H.~C., {et~al.} 2011, \apjs,
  197, 36, \dodoi{10.1088/0067-0049/197/2/36}

\bibitem[{{Lagos} {et~al.}(2018){Lagos}, {Stevens}, {Bower}, {Davis},
  {Contreras}, {Padilla}, {Obreschkow}, {Croton}, {Trayford}, {Welker}, \&
  {Theuns}}]{1701.04407}
{Lagos}, C. d.~P., {Stevens}, A. R.~H., {Bower}, R.~G., {et~al.} 2018, \mnras,
  473, 4956, \dodoi{10.1093/mnras/stx2667}

\bibitem[{{L'Huillier} {et~al.}(2012){L'Huillier}, {Combes}, \&
  {Semelin}}]{L'Huillier2012}
{L'Huillier}, B., {Combes}, F., \& {Semelin}, B. 2012, \aap, 544, A68,
  \dodoi{10.1051/0004-6361/201117924}

\bibitem[{{Lintott} {et~al.}(2011){Lintott}, {Schawinski}, {Bamford}, {Slosar},
  {Land}, {Thomas}, {Edmondson}, {Masters}, {Nichol}, {Raddick}, {Szalay},
  {Andreescu}, {Murray}, \& {Vandenberg}}]{Lintott2011}
{Lintott}, C., {Schawinski}, K., {Bamford}, S., {et~al.} 2011, \mnras, 410,
  166, \dodoi{10.1111/j.1365-2966.2010.17432.x}

\bibitem[{Lotz {et~al.}(2004)Lotz, Primack, \& Madau}]{Lotz2004}
Lotz, J.~M., Primack, J., \& Madau, P. 2004, The Astronomical Journal, 128,
  163, \dodoi{10.1086/421849}

\bibitem[{Lotz {et~al.}(2008)Lotz, Davis, Faber, Guhathakurta, Gwyn, Huang,
  Koo, Le~Floc’h, Lin, Newman, Noeske, Papovich, Willmer, Coil, Conselice,
  Cooper, Hopkins, Metevier, Primack, Rieke, \& Weiner}]{Lotz2008}
Lotz, J.~M., Davis, M., Faber, S.~M., {et~al.} 2008, The Astrophysical Journal,
  672, 177, \dodoi{10.1086/523659}

\bibitem[{{Lovell} {et~al.}(2021){Lovell}, {Vijayan}, {Thomas}, {Wilkins},
  {Barnes}, {Irodotou}, \& {Roper}}]{Lovell2021}
{Lovell}, C.~C., {Vijayan}, A.~P., {Thomas}, P.~A., {et~al.} 2021, \mnras, 500,
  2127, \dodoi{10.1093/mnras/staa3360}

\bibitem[{Marshall {et~al.}(2020)Marshall, Ni, Matteo, Wyithe, Wilkins, Croft,
  \& Kuusisto}]{Marshall2020}
Marshall, M.~A., Ni, Y., Matteo, T.~D., {et~al.} 2020, MNRAS, 499, 3819,
  \dodoi{10.1093/mnras/staa2982}

\bibitem[{{Marshall} {et~al.}(2022{\natexlab{a}}){Marshall}, {Watts},
  {Wilkins}, {Matteo}, {Kuusisto}, {Roper}, {Vijayan}, {Ni}, {Feng}, \&
  {Croft}}]{Madeline2022}
{Marshall}, M.~A., {Watts}, K., {Wilkins}, S., {et~al.} 2022{\natexlab{a}},
  \mnras, 516, 1047, \dodoi{10.1093/mnras/stac2111}

\bibitem[{{Marshall} {et~al.}(2022{\natexlab{b}}){Marshall}, Watts, Wilkins,
  Matteo, Kuusisto, Roper, Vijayan, Ni, Feng, \& Croft}]{MockCatalogue}
{Marshall}, M.~A., Watts, K., Wilkins, S., {et~al.} 2022{\natexlab{b}},
  BlueTides Mock Image Catalogue,  STScI/MAST, \dodoi{10.17909/ER09-4527}

\bibitem[{Mo {et~al.}(2010)Mo, van~den Bosch, \& White}]{Mo2010}
Mo, H., van~den Bosch, F.~C., \& White, S. 2010, {Galaxy Formation and
  Evolution}

\bibitem[{{Monachesi} {et~al.}(2019){Monachesi}, {G{\'o}mez}, {Grand},
  {Simpson}, {Kauffmann}, {Bustamante}, {Marinacci}, {Pakmor}, {Springel},
  {Frenk}, {White}, \& {Tissera}}]{Monachesi2019}
{Monachesi}, A., {G{\'o}mez}, F.~A., {Grand}, R. J.~J., {et~al.} 2019, \mnras,
  485, 2589, \dodoi{10.1093/mnras/stz538}

\bibitem[{Mortlock {et~al.}(2013)Mortlock, Conselice, Hartley, Ownsworth, Lani,
  Bluck, Almaini, Duncan, Wel, Koekemoer, Dekel, Dave, Ferguson, de~Mello,
  Newman, Faber, Grogin, Kocevski, \& Lai}]{Mortlock2013}
Mortlock, A., Conselice, C.~J., Hartley, W.~G., {et~al.} 2013, Monthly Notices
  of the Royal Astronomical Society, 433, 1185, \dodoi{10.1093/mnras/stt793}

\bibitem[{{Nelson} {et~al.}(2019){Nelson}, {Springel}, {Pillepich},
  {Rodriguez-Gomez}, {Torrey}, {Genel}, {Vogelsberger}, {Pakmor}, {Marinacci},
  {Weinberger}, {Kelley}, {Lovell}, {Diemer}, \& {Hernquist}}]{Nelson2019}
{Nelson}, D., {Springel}, V., {Pillepich}, A., {et~al.} 2019, Computational
  Astrophysics and Cosmology, 6, 2, \dodoi{10.1186/s40668-019-0028-x}

\bibitem[{{Nelson} {et~al.}(2022){Nelson}, {Suess}, {Bezanson}, {Price}, {van
  Dokkum}, {Leja}, {Whitaker}, {Labb{\'e}}, {Barrufet}, {Brammer},
  {Eisenstein}, {Heintz}, {Johnson}, {Mathews}, {Miller}, {Oesch}, {Sandles},
  {Setton}, {Speagle}, {Tacchella}, {Tadaki}, \& {Weaver}}]{NELSON2022}
{Nelson}, E.~J., {Suess}, K.~A., {Bezanson}, R., {et~al.} 2022, arXiv e-prints,
  arXiv:2208.01630.
\newblock \doarXiv{2208.01630}

\bibitem[{{Oesch} {et~al.}(2010){Oesch}, {Bouwens}, {Carollo}, {Illingworth},
  {Trenti}, {Stiavelli}, {Magee}, {Labb{\'e}}, \& {Franx}}]{Oesch2010}
{Oesch}, P.~A., {Bouwens}, R.~J., {Carollo}, C.~M., {et~al.} 2010, \apjl, 709,
  L21, \dodoi{10.1088/2041-8205/709/1/L21}

\bibitem[{Park {et~al.}(2022)Park, Lee, Kim, Jeong, Pichon, Gibson, Snaith,
  Shin, Kim, Dubois, \& Few}]{Park2022}
Park, C., Lee, J., Kim, J., {et~al.} 2022, The Astrophysical Journal, 937, 15,
  \dodoi{10.3847/1538-4357/ac85b5}

\bibitem[{Perrin {et~al.}(2015)Perrin, Long, Sivaramakrishnan, Lajoie, Elliot,
  Pueyo, \& Albert}]{Perrin2015}
Perrin, M.~D., Long, J., Sivaramakrishnan, A., {et~al.} 2015, WebbPSF: James
  Webb Space Telescope PSF Simulation Tool.
\newblock \doeprint{1504.007}

\bibitem[{{Petrosian}(1976)}]{Petrosian1976}
{Petrosian}, V. 1976, \apjl, 210, L53, \dodoi{10.1086/182301}

\bibitem[{{Pillepich} {et~al.}(2015){Pillepich}, {Madau}, \&
  {Mayer}}]{Pillepich2015}
{Pillepich}, A., {Madau}, P., \& {Mayer}, L. 2015, \apj, 799, 184,
  \dodoi{10.1088/0004-637X/799/2/184}

\bibitem[{{Reback} {et~al.}(2022){Reback}, {jbrockmendel}, {McKinney}, {Van den
  Bossche}, {Augspurger}, {Roeschke}, {Hawkins}, {Cloud}, {gfyoung}, {Sinhrks},
  {Hoefler}, {Klein}, {Petersen}, {Tratner}, {She}, {Ayd}, {Naveh},
  {Darbyshire}, {Garcia}, {Shadrach}, {Schendel}, {Hayden}, {Saxton},
  {Gorelli}, {Li}, {Zeitlin}, {Jancauskas}, {McMaster}, {W{\"o}rtwein}, \&
  {Battiston}}]{pandas}
{Reback}, J., {jbrockmendel}, {McKinney}, W., {et~al.} 2022,
  {pandas-dev/pandas: Pandas 1.4.2}, v1.4.2, Zenodo,  Zenodo,
  \dodoi{10.5281/zenodo.3509134}

\bibitem[{{Robertson} {et~al.}(2022){Robertson}, {Tacchella}, {Johnson},
  {Hausen}, {Alabi}, {Boyett}, {Bunker}, {Carniani}, {Egami}, {Eisenstein},
  {Hainline}, {Helton}, {Ji}, {Kumari}, {Lyu}, {Maiolino}, {Nelson}, {Rieke},
  {Shivaei}, {Sun}, {Ubler}, {Williams}, {Willmer}, \& {Witstok}}]{MORFEUS2022}
{Robertson}, B.~E., {Tacchella}, S., {Johnson}, B.~D., {et~al.} 2022, arXiv
  e-prints, arXiv:2208.11456.
\newblock \doarXiv{2208.11456}

\bibitem[{Rodriguez-Gomez {et~al.}(2015)Rodriguez-Gomez, Genel, Vogelsberger,
  Sijacki, Pillepich, Sales, Torrey, Snyder, Nelson, Springel, Ma, \&
  Hernquist}]{Rodriguez-Gomez2015}
Rodriguez-Gomez, V., Genel, S., Vogelsberger, M., {et~al.} 2015, Monthly
  Notices of the Royal Astronomical Society, 449, 49,
  \dodoi{10.1093/mnras/stv264}

\bibitem[{{Rodriguez-Gomez} {et~al.}(2022){Rodriguez-Gomez}, {Genel}, {Fall},
  {Pillepich}, {Huertas-Company}, {Nelson}, {P{\'e}rez-Monta{\~n}o},
  {Marinacci}, {Pakmor}, {Springel}, {Vogelsberger}, \&
  {Hernquist}}]{RodriguezGomez2022}
{Rodriguez-Gomez}, V., {Genel}, S., {Fall}, S.~M., {et~al.} 2022, \mnras, 512,
  5978, \dodoi{10.1093/mnras/stac806}

\bibitem[{{Roper} {et~al.}(2022){Roper}, {Lovell}, {Vijayan}, {Marshall},
  {Irodotou}, {Kuusisto}, {Thomas}, \& {Wilkins}}]{Roper2022}
{Roper}, W.~J., {Lovell}, C.~C., {Vijayan}, A.~P., {et~al.} 2022, \mnras, 514,
  1921, \dodoi{10.1093/mnras/stac1368}

\bibitem[{{Rose} {et~al.}(2022){Rose}, {Kartaltepe}, {Snyder},
  {Rodriguez-Gomez}, {Yung}, {Arrabal Haro}, {Bagley}, {Calabr{\`o}}, {Cleri},
  {Cooper}, {Costantin}, {Croton}, {Dickinson}, {Finkelstein},
  {H{\"a}u{\ss}ler}, {Holwerda}, {Koekemoer}, {Kurczynski}, {Lucas}, {Mantha},
  {Papovich}, {P{\'e}rez-Gonz{\'a}lez}, {Pirzkal}, {Somerville}, {Straughn}, \&
  {Tacchella}}]{Rose2022}
{Rose}, C., {Kartaltepe}, J.~S., {Snyder}, G.~F., {et~al.} 2022, arXiv
  e-prints, arXiv:2208.11164.
\newblock \doarXiv{2208.11164}

\bibitem[{{Rosse}(1850)}]{Rosse1850}
{Rosse}, T. E.~O. 1850, Philosophical Transactions of the Royal Society of
  London Series I, 140, 499

\bibitem[{{Schawinski} {et~al.}(2014){Schawinski}, {Urry}, {Simmons},
  {Fortson}, {Kaviraj}, {Keel}, {Lintott}, {Masters}, {Nichol}, {Sarzi},
  {Skibba}, {Treister}, {Willett}, {Wong}, \& {Yi}}]{Schawinski2014}
{Schawinski}, K., {Urry}, C.~M., {Simmons}, B.~D., {et~al.} 2014, \mnras, 440,
  889, \dodoi{10.1093/mnras/stu327}

\bibitem[{{Schaye} {et~al.}(2015){Schaye}, {Crain}, {Bower}, {Furlong},
  {Schaller}, {Theuns}, {Dalla Vecchia}, {Frenk}, {McCarthy}, {Helly},
  {Jenkins}, {Rosas-Guevara}, {White}, {Baes}, {Booth}, {Camps}, {Navarro},
  {Qu}, {Rahmati}, {Sawala}, {Thomas}, \& {Trayford}}]{Schaye2015}
{Schaye}, J., {Crain}, R.~A., {Bower}, R.~G., {et~al.} 2015, \mnras, 446, 521,
  \dodoi{10.1093/mnras/stu2058}

\bibitem[{S{\'{e}}rsic(1963)}]{Sersic1963}
S{\'{e}}rsic, J.~L. 1963, Boletin de la Asociacion Argentina de Astronomia, 6.
\newblock \url{http://adsabs.harvard.edu/abs/1963BAAA....6...41S}

\bibitem[{{Sparre} \& {Springel}(2017)}]{Sparre2017}
{Sparre}, M., \& {Springel}, V. 2017, \mnras, 470, 3946,
  \dodoi{10.1093/mnras/stx1516}

\bibitem[{{Stefanon} {et~al.}(2017){Stefanon}, {Yan}, {Mobasher}, {Barro},
  {Donley}, {Fontana}, {Hemmati}, {Koekemoer}, {Lee}, {Lee}, {Nayyeri}, {Peth},
  {Pforr}, {Salvato}, {Wiklind}, {Wuyts}, {Ashby}, {Castellano}, {Conselice},
  {Cooper}, {Cooray}, {Dolch}, {Ferguson}, {Galametz}, {Giavalisco}, {Guo},
  {Willner}, {Dickinson}, {Faber}, {Fazio}, {Gardner}, {Gawiser}, {Grazian},
  {Grogin}, {Kocevski}, {Koo}, {Lee}, {Lucas}, {McGrath}, {Nandra}, {Newman},
  \& {van der Wel}}]{Stefanon2017}
{Stefanon}, M., {Yan}, H., {Mobasher}, B., {et~al.} 2017, \apjs, 229, 32,
  \dodoi{10.3847/1538-4365/aa66cb}

\bibitem[{{Thob} {et~al.}(2019){Thob}, {Crain}, {McCarthy}, {Schaller},
  {Lagos}, {Schaye}, {Talens}, {James}, {Theuns}, \& {Bower}}]{Thob2019}
{Thob}, A. C.~R., {Crain}, R.~A., {McCarthy}, I.~G., {et~al.} 2019, \mnras,
  485, 972, \dodoi{10.1093/mnras/stz448}

\bibitem[{{Tissera} {et~al.}(2012){Tissera}, {White}, \&
  {Scannapieco}}]{Tissera2012}
{Tissera}, P.~B., {White}, S. D.~M., \& {Scannapieco}, C. 2012, \mnras, 420,
  255, \dodoi{10.1111/j.1365-2966.2011.20028.x}

\bibitem[{{Trayford} {et~al.}(2019){Trayford}, {Frenk}, {Theuns}, {Schaye}, \&
  {Correa}}]{Trayford2019}
{Trayford}, J.~W., {Frenk}, C.~S., {Theuns}, T., {Schaye}, J., \& {Correa}, C.
  2019, \mnras, 483, 744, \dodoi{10.1093/mnras/sty2860}

\bibitem[{{Whitney} {et~al.}(2021){Whitney}, {Ferreira}, {Conselice}, \&
  {Duncan}}]{Whitney2021}
{Whitney}, A., {Ferreira}, L., {Conselice}, C.~J., \& {Duncan}, K. 2021, \apj,
  919, 139, \dodoi{10.3847/1538-4357/ac1422}

\bibitem[{{Zana} {et~al.}(2022){Zana}, {Lupi}, {Bonetti}, {Dotti},
  {Rosas-Guevara}, {Izquierdo-Villalba}, {Bonoli}, {Hernquist}, \&
  {Nelson}}]{Zana2022}
{Zana}, T., {Lupi}, A., {Bonetti}, M., {et~al.} 2022, \mnras, 515, 1524,
  \dodoi{10.1093/mnras/stac1708}

\end{thebibliography}
\bibliographystyle{aasjournal}



\end{document}